\documentclass{article}

\usepackage{amssymb,amsmath,hyperref,graphicx,color,makeidx,chngpage}

\newcommand{\tr}{\textrm{Tr}}

\begin{document}
\title{\bf{Functional Renormalization of Noncommutative Scalar Field Theory}}
\author{Alessandro Sfondrini\\
Dipartimento di Fisica, Universit\`a di Padova,\\ and INFN, Sezione di Padova,\\
via Marzolo 8, 31131 Padova\\ Italy \\
{\it and}\\
 Tim A. Koslowski\\ \texttt{tkoslowski@perimeterinstitute.ca}\\
Perimeter Institute for Theoretical Physics,\\
31 Caroline Street North,\\ Waterloo, Ontario N2J 2Y5\\Canada\\}
\date{\today}
\maketitle

\begin{abstract}
In this paper we apply the Functional Renormalization Group Equation (FRGE)  to the non-commutative scalar field theory proposed by Grosse and Wulkenhaar. We derive the flow equation in the matrix representation and discuss the theory space for the self-dual model. The features introduced by the external dimensionful scale provided by the non-commutativity parameter, originally pointed out in \cite{Gurau:2009ni}, are discussed in the FRGE context. Using a technical assumption, but without resorting to any truncation, it is then shown that the theory is asymptotically safe for suitably small values of the $\phi^4$ coupling, recovering the result of \cite{disertori:2007}. Finally, we show how the FRGE can be easily used to compute the one loop beta-functions of the duality covariant model.
\end{abstract}

\section{Introduction}

The Functional Renormalization Group Equation (FRGE) is a powerful tool in the study of interacting Quantum Field Theories (QFT) and statistical systems. The FRGE  proposed in \cite{wetterich:1991,wetterich:1993}  describes the evolution of the quantum effective average action of a theory when changing the coarse graining scale, and has found a number of applications in statistical Physics \cite{bergestetradis:2002}, particle Physics and in the \lq\lq asymptotic safety\rq\rq\ conjecture in Quantum Gravity \cite{niedermaier:2006,codellopercacci:2009}. Despite appearing as a one-loop equation, it accounts for arbitrarily many-loop effects through effective operators, and it is particularly suited for the study of non-perturbative questions\footnote{The term non-perturbative is here referred to systems that do not admit a treatment in perturbation theory, and should not be misunderstood as the inequivalent notion of non-perturbative in constructive field theory.}. In what follows, we refer to approaches relying on the flow of the effective average action as \lq\lq functional renormalization\rq\rq, and call approaches that study the running of the action \lq\lq Wilsonian\rq\rq.

Technically, the FRGE is a differential equation defined on a \emph{theory space}, i.e. on the infinite dimensional space of action functionals compatible with the field content and symmetries of the QFT we want to consider. In practice, this concept often boils down to considering a truncation ansatz for the theory space, which is usually much \lq\lq smaller\rq\rq\ than that or even finite-dimensional. Most truncations are approximations as they are \emph{not stable under the RG flow}. It is thus interesting in its own right to study the flow of a physical theory without resorting to any truncation. This would e.g. lead to understanding whether the ultraviolet (UV) behavior of Quantum Gravity is dominated by an attractive fixed point with a finite number of relevant directions as suggested by the asymptotic safety conjecture. Unfortunately, such an approach presents several technical difficulties, which appear very hard to overcome. This provides the motivation for this paper: we consider a much simpler model that can be studied on an entire theory space.

Non-commutative field theories may appear an awkward choice for a testing ground for the FRGE. Even if they are physically interesting, as they arise naturally as a limiting regime of some string theories \cite{connes:1998} and may provide a description of the quantum Hall effect \cite{douglas:2001}, their renormalization is rather problematic. The presence of a fundamental length scale, encoded in the dimensionful non-commutativity parameter $\theta$, may suggest that the theory is well behaved in the UV. However, it turns out that in absence of an UV cutoff many theories have several infrared (IR) divergent graphs that, when inserted in larger graphs, spoil their renormalizability: this is the \lq\lq UV/IR mixing\rq\rq\ problem \cite{minwalla:2000}.

A way around this problem has been found by Grosse and Wulkenhaar (GW), who proved that a particular, $\phi^4$-like, scalar field theory on Moyal space is perturbatively renormalizable \cite{grossewulkenhaar:2003,grossewulkenhaar:2005a,grossewulkenhaar:2005b,Gurau:2005gd}. Before that, Langmann and Szabo already noticed that such a model has a remarkable covariance under the interchange of positions and momenta \cite{langmannszabo:2002} which becomes invariance for a particular choice of the bare action. We call this the self-dual GW model. This, together with the fact that the GW model can be formulated as a matrix theory, which is particularly simple in self-dual case, allowed to prove that \emph{the $\beta$-function of the model vanishes to all orders of perturbation theory} \cite{disertori:2007}. Thus, the renormalization group flow of the $\phi^4$-coupling constant is bounded, and the theory is \emph{asymptotically safe}. This is much better than ordinary $\phi^4$ theory, which appears to be plagued by the infamous Landau ghost.

In this paper we  use the matrix formulation of the GW model and study its renormalization with the FRGE. The second motivation for this paper is to show how convenient FRGE techniques are to study matrix theories. To do so, we will adapt the FRGE formalism to matrix theories, and in particular discuss the features introduced by the dimensionful scale $\theta$. We will use slightly different reasoning compared to what was used in Wilsonian renormalization in \cite{Gurau:2009ni}, but ultimately reach equivalent conclusions.

The paper is structured as follows. First, we introduce the Moyal space as a setting for non-commutative QFTs, present the associated matrix representation, and formulate the GW model in both languages. Second, we recover the flow equation in the matrix base, with attention to the definition of the theory space for the self-dual theory. As mentioned, we discuss the fixed point condition and asymptotic safety in presence of a fundamental scale. In the following section, we explicitly compute the one-loop $\beta$-functions of the duality invariant model, which we compare with \cite{grossewulkenhaar:2004}, and use them to show \emph{without resorting to any truncation} that, under some technical assumptions, the model is asymptotically safe. We show in particular that there is a one-dimensional attractor in the vicinity of the Gaussian fixed point, essentially given by a finite $\phi^4$-coupling. Finally, we show how the computation can be extended to include duality-covariant terms, in a simple truncation, and easily recover the perturbative result that the self-dual theory is UV attractive at one-loop \cite{grossewulkenhaar:2004}.

\section{Setup}

The Grosse-Wulkenhaar model is a bare action for a scalar field theory on a noncommutative deformation of $\mathbb R^4$, following \cite{gracia-bondia:1988,grossewulkenhaar:2003}. We will now introduce the effective average action setup for this model: the noncommutative deformation (Section \ref{sec:nc-deformation}),  the FRGE (Sec. \ref{sec:wetterich-equation}) and examine the self-dual theory space (Sec. \ref{sec:theory-space}) as well as discussing \lq\lq fundamentality\rq\rq\ in this theory space (Sec. \ref{sec:fundamental}).

\subsection{Noncommutative Space and Matrix Representation}\label{sec:nc-deformation}

For $p\in \mathbb R^4$ we define the symbol $e_p$ with product $e_p  e_q=e_{p+q} e^{\frac i 2 p^T.\Theta.q}$,
where $\Theta^{\mu\nu}=\theta\left(\varepsilon^{\mu\nu34}+\varepsilon^{12\mu\nu}\right)$.\footnote{We take Levi-Civita symbol to yield $\varepsilon^{1234}=1$.} This requires the introduction of a constant $\theta$, which has the dimensions of a length squared. Defining the evaluation $e_p(x):=e^{ix^T.p}$ for $x\in \mathbb R^4$ we consider the associative algebra over $\mathbb C$ of Schwartz functions $f\in \mathcal S(\mathbb R^4)$ with product
\begin{equation}
  f_1 \star f_2 := \int d^4p d^4 q \tilde f_1(p)\tilde f_2(q) e_p e_q,
\end{equation}
where $\tilde f(p):=\int \frac{d^4x}{(2\pi)^4}f(x)e^{i x^T.p}$. 
We call this the \emph{position representation}. Let us now introduce a matrix representation of this algebra: using the idempotent $f_o(x)=4 e^{-\frac{x^T.x}\theta}$ and the harmonic oscillator ladder operators $a_1=\frac{1}{\sqrt{2}}\left(x_1+i x_2\right)$ and $a_2=\frac{1}{\sqrt{2}}\left(x_3+i x_4\right)$, we define the complete set of linearly independent functions
\begin{equation}
  f_{n_1m_1n_2m_2}=\frac{(a_1^*)^{\star m_1}(a_2^*)^{\star m_2}\star f_o\star (a_1)^{\star n_1}(a_2)^{\star n_2}}{\sqrt{\theta^{n_1+n_2+m_1+m_2}n_1!n_2!m_1!m_2!}}\;,
\end{equation}
such that $\int d^4 x f_{n_1m_1n_2m_2}=(2\pi \theta)^2 \delta_{n_1m_1}\delta_{n_2m_2}$. This allows us to construct the matrix representation by expanding
\begin{equation}
  \phi(x)=\sum_{n_1m_1n_2m_2} \phi_{n_1m_1n_2m_2} f_{n_1m_1n_2m_2}(x)\;,
\end{equation}
which remarkably turns  the $\star$-product into a matrix product
\begin{equation}
  (\phi\star \psi)_{n_1m_1n_2m_2}=\sum_{k_1k_2} \phi_{n_1k_1n_2k_2}\psi_{k_1m_1k_2m_2}\;,
\end{equation}
and the volume integral into a matrix trace
\begin{equation}
  \int d^4 x \phi(x)=(2\pi\theta)^2 \sum_{n_1n_2} \phi_{n_1n_1n_2n_2}\;.
\end{equation}
To simplify notation we will henceforth condense the quadruple $(n_1,m_1,n_2,m_2)$ into a pair of matrix indices $(n,m)$, and indicate the matrix product with a dot.  A general $\star$-local functional $F$ of a field $\phi$ on $\mathbb R^4_\theta$ can be expanded in monomials both in the matrix representation and in the position representation
\begin{equation}
  F[\phi]=\sum_{i,\vec j_i} \tr\left(\phi.A_{j_1}\,...\,\phi.A_{j_{n_i}}\right)=\sum_{i,\vec j_i} \int d^4 x \phi(x)\star A_{j_1}(x)\star ...\star \phi\star A_{j_{n_i}}(x)\;.
\end{equation}
In particular, the ladder operators in the matrix representation have matrix elements
\begin{equation}\label{equ:ladder-matrix}
(a_1)_{m n}=\sqrt{(m+1)\theta}\;\delta_{m_1+1,n_1}\delta_{m_2,n_2}, \ \ \ (a_1^*)_{m n}=\sqrt{m\theta}\;\delta_{m_1,n_1+1}\delta_{m_2,n_2}\;,
\end{equation}
and likewise for $a_2,a_2^*$. By construction, these operators obey canonical commutation relations upon normalization.

\subsubsection*{The Grosse -Wulkenhaar Model}

Using $\tilde x_\mu:=2 (\theta^{-1})_{\mu\nu}x^\nu$, the action proposed by Grosse and Wulkenhaar \cite{grossewulkenhaar:2003} is
\begin{align}\nonumber
  S[\phi(x)]=\int d^4 x &\left(\frac 1 2 (\partial_\mu\phi)\star(\partial^\mu \phi)+\frac{1}{2}\Omega^2(\tilde x_\mu\phi)\star(\tilde x^\mu \phi)+\right.\\
  \label{equ:GW-action-position}
  &\ \ \ \left.+\frac {1}{2}m^2 \phi\star\phi+\frac{1}{4!}\lambda \phi\star\phi\star\phi\star\phi\right)\;,
\end{align}
where $\Omega>0$ is a dimensionless parameter. The field $\phi(x)$ has the dimension of a mass, as in the commutative case. This model breaks translation invariance, as the second term adds an harmonic oscillator potential. Under  the interchange of position and momenta $p_\mu \leftrightarrow \tilde x_\mu$ and $\hat\phi(p) \leftrightarrow \pi^2 \theta^2 \int d^4 x e^{(-1)^j p^\mu x_\mu} \phi(x)$ (where $j$ denotes the position of $\phi$ in a monomial $\phi\star ...\star \phi$) the Grosse-Wulkenhaar action transforms covariantly:
\begin{equation}
  S(\Omega,m,\lambda) \to \Omega^2 \hat{S}(\frac{1}{\Omega},\frac{m}{\Omega},\frac{\lambda}{\Omega^2}).
\end{equation}
The action is self-dual when $\Omega=1$. The symmetry of the self-dual action is particularly simple in the matrix representation. Introducing four matrix operators
\begin{equation}
\tilde{X}_{2k-1}:=\frac{a^*_k+a_k}{\sqrt{2}\theta}, \ \ \ \ \ \tilde{X}_{2k}:=\frac{a^*_k-a_k}{i\sqrt{2}\theta}, \ \ \ \ \ k=1,2\;,
\end{equation}
and defining $\omega:=\Omega^2-1$ and the volume element $\nu:=(2\pi\theta)^2$,  the action (\ref{equ:GW-action-position})  reads in the matrix basis
\begin{align}\nonumber
  S[\phi_{mn}]=\nu \tr&\left(\frac{1+\frac \omega 2}{2} \phi.\tilde X_\mu.\tilde X^\mu.\phi+\frac{1}{2} \omega\; \phi.\tilde X_\mu.\phi.\tilde X^\mu +\right.\\
 \label{equ:GW-action-matrix}
  &\ \ \ \ \left.+\frac{1}{2}m^2\; \phi.\phi + \frac{1}{4!} \lambda\;\phi.\phi.\phi.\phi\right)\;,
\end{align}
and $\phi_{mn}$ retains the same mass dimension as in the real space. Using the explicit form of $\tilde X_\mu$, at $\Omega=1$ ($\omega=0$) the diagonal kinetic term is
\begin{equation}
  \phi.\tilde X_\mu.\tilde X^\mu.\phi = \phi_{mn}\mathcal{K}_{mnkl}\phi_{kl}=\phi_{mn} \frac 4 \theta(n_1+m_1+n_2+m_2+2)\delta_{ml}\delta_{nk}\phi_{kl},
\end{equation}
which yields propagator by straightforward inversion. The term $\phi.\tilde X_\mu.\phi.\tilde X^\mu$ would give a non-diagonal matrix operator, whose inversion is possible \cite{grossewulkenhaar:2005b}, but rather cumbersome. It will also be useful to introduce the matrix multiplication by the operator $K_{mn}:=\frac{4}{\theta}(m_1+m_2+1)\delta_{mn}$, rather than working with the four-indices Hessian $\mathcal{K}_{mnkl}$, so that
\begin{equation}
\tr\left[\phi.\tilde X_\mu.\tilde X^\mu.\psi\right]=\sum_{m,n,k,l}\phi_{mn}\mathcal{K}_{mnkl}\psi_{kl}=\tr\left[\phi.K.\psi\right]+\tr\left[\psi.K.\phi\right]\;.
\end{equation}

\subsection{Flow Equation for the Effective Average Action}\label{sec:wetterich-equation}

To derive an RG flow equation in matrix base and fix notations, we follow Wetterich's original approach \cite{wetterich:1991,wetterich:1993} and consider the partition function
\begin{equation}\label{equ:partition-function}
  e^{W_k[j]}=\int D_\Lambda \chi \exp\left(-S_\Lambda[\chi]-\Delta_k S[\chi]+\nu \tr(j^T.\chi) \right),
\end{equation}
with an overall sharp cutoff in the UV (i.e. in matrix size)
\begin{equation}\label{equ:cutoff-measure}
  D_\Lambda \chi = \prod_{n=0}^{\Lambda^2\theta}\prod_{m=0}^{\Lambda^2\theta} d\chi_{mn}.
\end{equation}
so that the functional measure is a well-defined finite product of Lebesgue measures. We add a quadratic IR suppression term $\Delta_k S[\chi]=\frac 1 2 \chi_{ab}R_{k}^{abcd}\chi_{cd}$\footnote{For convenience, we incorporate $\nu$ in $R_k$, which has now mass dimension $-2$. Furthermore, when this does not generate confusion, we suppress the subscript index $k$ in $R_k$, $W_k$, etc.} that modifies the spectrum of the kinetic operator $\mathcal{K}$ for eigenvalues smaller than $k^2$ (the IR modes) by adding a positive term of order $k^2$, and leaves remaining modes (the UV) unchanged.
Let us introduce
\begin{equation}
  \begin{array}{rcl}
    \phi_{ab}&:=&\langle \chi_{ab}\rangle=\nu^{-1} \left.\frac{\delta}{\delta j_{ab}}e^{W_k[j]}\right|_{j=0}\;,\\
    \label{equ:correlation}
    C_{abcd}&:=&\langle \chi_{ab}\chi_{cd}\rangle=\phi_{ab}\phi_{cd}+\left.\nu^{-2} \frac{\delta^2 W_k[j]}{\delta j_{ab}\delta j_{cd}}\right|_{j=0}\;.
  \end{array}
\end{equation}
We introduce the partial and total scale derivatives $\partial_t:=k\partial_k$ and $d_t$ respectively (the latter will be also indicated by a dot). We have from (\ref{equ:partition-function})  that
\begin{equation}\label{equ:W-derivative}
  \left.\partial_t W_k\right|_{j} = -\frac{1}{2}C_{abcd}\dot{R}_k^{abcd}\;.
\end{equation}
We require that $R_k^{abcd}$ vanishes (or decays fast enough) for $a,b\gtrsim k^2\theta$ , as well as a fast enough decay of $d_t  R_k^{abcd}$.
The effective average action is
  \begin{align}
\nonumber
    \Gamma_k[\phi]&:=\sup_j\left\{\nu\, \tr(j^T.\phi)-W_k[j]\right\} -\Delta_k S[\phi]\\
\label{equ:effective-average-action}
      &\;=\nu\, \tr(j_{k*}^T.\phi)-W_k[j_{k*}]-\Delta_k S[\phi],
  \end{align}
where $j_{k*}$ denotes the (by assumption) unique $j_k$ where the supremum is attained. The total scale derivative of the effective average action is, using (\ref{equ:correlation}, \ref{equ:W-derivative})
\begin{align}
\nonumber
   \dot{\Gamma}_k[\phi]&=\nu\, \tr\left(\phi.d_t j_{k*}^T\right)-\left.\partial_t W_k[j]\right|_{j=j_{k*}}\!\!\!\!-\left.\frac{\delta W_k[j]}{\delta j_{rs}}\right|_{j=j_{k*}}\!\!\!\!\!\!\!\!d_t j_{k*rs}-\frac 1 2\,\phi_{ab}\phi_{cd}d_t R_k^{abcd}\\
     &=- \frac 1 2 \left.\nu^{-2}\frac{\delta^2 W_k[j]}{\delta j_{ab}\delta j_{cd}}\right|_{j=j_{k*}}\!\!\!\! \dot{R}_k^{abcd}\;.
\end{align}
Since $\frac{\delta\phi_{rs}}{\delta\phi_{tu}}=\delta_{rt}\delta_{su}$, which is the identity in matrix space, we get the familiar formula for the two point function
\begin{equation}
\left(\Gamma^{(2)} + R \right)_{xyrs}^{-1}[\phi]=-\nu^{-2}\left. \frac{\delta^2 W_k[j]}{\delta j_{xy}\delta j_{rs}}\right|_{j=j_k*[\phi]},
\end{equation}
where we used an obvious short-hand for functional derivatives. The FRGE takes the form
\begin{equation}\label{equ:wetterich-equation}
  \dot{\Gamma}_k[\phi]=\frac 1 2 (\Gamma^{(2)}+R)^{-1}_{abcd}\dot{R}_k^{abcd}\;,
\end{equation}
where $R_k$ provides both an IR and a UV cutoff (via its scale derivative) at $a,b\approx k^2\theta$. In what follows we evaluate this equation in a vertex expansion. Let
\begin{equation}
  \frac{\delta^2(\Gamma_k+\Delta_k S)[\phi]}{\delta\phi_{tu}\delta\phi_{ab}}=G_{tuab}+F_{tuab}[\phi]\;,
\end{equation}
by construction, the field independent part is IR regulated and can be inverted. Defining $P:=G^{-1}$ and using $(\Gamma^{(2)}+R)^{-1}=\sum_{n=0}^\infty (-PF)^nP$, the vertex expansion for (\ref{equ:wetterich-equation}) is
\begin{equation}\label{equ:vertex-expansion}
  \dot{\Gamma}_k[\phi]=\frac 1 2 \left(P_{abcd}-P_{abrs}F_{rstu}P_{tucd}+...\right)\,\dot{R}_k^{abcd}\;.
\end{equation}
The most general ansatz for the vertex expansion of the effective average action compatible with the symmetry $\phi \to -\phi$ is of the form
\begin{equation}\label{equ:general-Gamma-k}
  \Gamma_k[\phi]=\sum_{i=1}^\infty \Gamma_k^{n_{1}m_{1}...n_{2i}m_{2i}}\phi_{n_1m_1}...\phi_{n_{2i}m_{2i}\;}\;,
\end{equation}
where  $m_{2i}$, $n_{2i}$ ect. are actually double indices, i.e. $m_{2i}\equiv\left([m_{2i}]_1,[m_{2i}]_2\right)$. The general Hessian takes the form
\begin{equation}\label{equ:general-Hessian}
  \Gamma^{(2)ab,cd}_k[\phi]=\sum_{i=1}^\infty F_k^{abcdn_1m_1...n_{2i-2}m_{2i-2}}\phi_{n_1m_1}...\phi_{n_{2i-2}m_{2i-2}}\;,
\end{equation}
where $F_k^{abcdn_1m_1...n_{2i-2}m_{2i-2}}$ is the sum over all permutations of multi-index pairs in $\Gamma_k^{n_{1}m_{1}...n_{2i}m_{2i}}$. The explicit form of the vertex expansion can be worked out by a tedious but straightforward computation once the theory space has been defined. In Appendix \ref{app:explicit-vertex} we explicitly provide the terms that we need to investigate asymptotic safety.

\subsubsection*{Choice of Regulators}

The partition function (\ref{equ:partition-function}) involves an overall UV cut-off measure (\ref{equ:cutoff-measure}).Thus, for $\Lambda_1>\Lambda_2$, we must impose the Wilsonian consistency condition
\begin{equation}
  e^{-S_{\Lambda_2}[\phi]}=\int \prod_{\Lambda_2\sqrt{\theta}<n,m\le \Lambda_1\sqrt{\theta}} d\phi_{nm} e^{-S_{\Lambda_1}[\phi]}\;,
\end{equation}
to ensure that $W_k$ in (\ref{equ:partition-function}) is independent of the UV cutoff $\Lambda\geq k$, for any fixed $k$. However, in (\ref{equ:wetterich-equation}) the UV cutoff is provided by the scale derivative of the regulator  $\dot{R}_k$ so that, if we choose $R_k$ appropriately, no mode corresponding to a scale larger than $k$ has to be summed over to obtain the flow equation for $\Gamma_k$. Then, the latter depends on the UV cutoff just formally, as long as we take $\Lambda\geq k$; in particular, as we will discuss in next section, it will be convenient to set $\Lambda\approx k$ after computing the flow equation, in order to identify the couplings. However, taking the limit $k\to\infty$ still requires some care, as we shall discuss in Section \ref{sec:fundamental}. Notice that the insensitivity of the flow equation to the cutoff $\Lambda$ is due to the fact that we work with the effective average action $\Gamma_k$, rather than directly with the Wilsonian action $S_{k,\Lambda}$.

For the self-dual model, we can require the regulator to be of the form
\begin{equation}\label{equ:generic-regulator}
\left(R_k\right)_{abcd}=\nu\,k^2\,\mathcal{Z}\; r\!\left(\frac{a}{\theta},\frac{b}{\theta}\right)\;\delta_{da}\delta_{bc}\;H\left[k^2\theta -\left\|a,b\right\|\right]\;,
\end{equation}
we have explicitly written the volume element $\nu$, the scaling dimension $k^2$ and a $k$-dependent dimensionless wave function renormalization $\mathcal Z=\mathcal Z(k)$. This is useful because the regulator appears in the two-point function. $H$ is Heaviside's step function. Observe that $R_k$ is diagonal in the matrix indices. The cutoff on the UV modes depends on the choice of the norm on the indices; a convenient choice is e.g. $\left\|(a,b)\right\|=\sup(a,b)$. To ensure that the cutoff is effective in the IR, we also have to require that $r(x,y)>0$ when $x,y\to0$, and that the function $r$ is monotonic. The physical predictions will then be largely independent of the details of the $r$ and of the norm, as expected from a similar feature in commutative field theories.

To carry out explicit calculations it is necessary to pick a particular $R_k$. We consider two choices. The first is constructed to model the matrix-size cut-off used in perturbation theory.
\begin{equation}\label{equ:regulator-sharp}
R_k^{abcd}=\nu\mathcal{Z}\;\alpha k^2\;\delta^{da}\delta^{bc} H\!\left[{\textstyle \beta k^2-\frac{a+1}{\theta}}\right]\,H\!\left[{\textstyle \beta k^2-\frac{b+1}{\theta}}\right]\;,
\end{equation}
with $\alpha>0$ and $\beta>0$ free parameters. It implements a mass term of order $\alpha k^2$ on the low momentum modes, i.e. on the modes such that $a_1+a_2+1\leq\beta\theta k^2$ and $b_1+b_2+1\leq\beta\theta k^2$. Its scale derivative is
\begin{equation}
\dot{R}_k^{abcd}=\nu\mathcal{Z}\;(1+\frac{\eta}{2})2\alpha k^2\;\delta_{ad}\delta_{bc} H\!\left[{\textstyle \beta k^2-\frac{a+1}{\theta}}\right]\,H\!\left[{\textstyle \beta k^2-\frac{b+1}{\theta}}\right]+...,
\end{equation}
where $\eta=\frac 1{\mathcal Z} \partial \mathcal Z$. This regulator provides an UV cutoff on the matrix size at order $\beta\theta k^2$ when inserted in the trace in (\ref{equ:vertex-expansion}). Furthermore $\dot{R}$ will also contain $\delta$-functions coming from e.g. $d_t H\!\left[{\textstyle \beta k^2-\frac{a+1}{\theta}}\right]$ that we omitted, since we will use this regulator only to mimic perturbative calculations and as such in truncations where these $\delta$-functions are either projected out or appear in terms that are suppressed for $k\to \infty$. The main difference between this regulator and the regularization scheme employed in perturbation theory, as in \cite{grossewulkenhaar:2004}, is the presence of the IR regulator $\alpha k^2$; as we will see, to recover the perturbative result we will have to formally send $\alpha\to0$ at the end of our calculation, and we will confirm that the result is independent of $\beta$.

The second choice for the regulator is to mimic Litim's \lq\lq optimized\rq\rq\ regulator \cite{litim:2001}
\begin{equation}\label{equ:regulator-litim}
R_k^{abcd}=\nu\mathcal{Z}\;\beta \left(k^2-{\textstyle \frac{a+b+2}{\beta\; \theta}}\right)\delta^{da}\delta^{bc} H\!\left[{\textstyle \beta k^2-\frac{a+1}{\theta}}\right]H\!\left[{\textstyle \beta k^2-\frac{b+1}{\theta}}\right].
\end{equation}
Here we have only one free parameter $\beta>0$, because we require $R$ to cancel the kinetic term coming from $\Gamma_k^{(2)}$. The derivative yields
\begin{equation}\label{equ:rdot-litim}
\dot{R}_k^{abcd}=\eta R_k^{abcd}+\nu\mathcal{Z}\;\beta 2k^2\delta^{da}\delta^{bc} H\!\left[{\textstyle \beta k^2-\frac{a+1}{\theta}}\right]H\!\left[{\textstyle \beta k^2-\frac{b+1}{\theta}}\right],
\end{equation}
where this time no $\delta$-functions are generated. A nice feature of this regulator is that in the truncation $\Gamma_k=S_k$ (see Section \ref{sec:oneloop-dual}), we have that $G=S^{(2)}|_{\phi=0}+R$ is a scalar matrix for $a,b$ smaller than the UV cutoff, so that its inversion is straightforward. This choice converges fast in a derivative expansion \cite{litim:2001,litim:2000} and reduces scheme effects. We will observe a similar feature later.

\subsubsection*{Identification of Field Monomials}

The vertex and derivative expansion are very powerful tools for the evaluation of traces appearing in exact RG equations. The derivative expansion is well adapted to commutative field theories, because (1) it is adapted to the principle of locality, (2) the dimension of the lowest dimensional operator at each order of the derivative expansion increases with the order of the expansion and (3) it allows to evaluate the traces successively by expanding the field configuration around homogeneous configurations. It is however poorly adapted to noncommutative field theories, since their position representation is nonlocal.

To construct a generalization of the derivative expansion that is suitable to the noncommutative setting, we observe that the noncommutative analogue of local products is the matrix product (in matrix representation), and the analogue of an integral is replaced by the matrix trace. Moreover, we observe that the ladder operators play a role analogous to the derivative operators. However, insertion of matrices with vanishing elements except in the upper left $n\times n$ entries, which is the analogue of an expansion around homogeneous fields, is not annihilated by any power of $K$. Thus, we can not use matrices with upper left $n\times n$ entries to project onto a truncation with a small number of $K$ operators\footnote{The $n$th power of a ladder operator does indeed annihilate a field matrix with only upper left $n\times n$ entries, but the combinations $(a^* a)^k$ do not.}. We thus have to adopt a different strategy to identify field monomials in the flow equation.

Let us order the field monomials by their dimension, so for any $d$ there is only a finite number of couplings with dimension smaller or equal $d$. We can then find a finite $n(d)$, and in fact a minimal $n_o(d)$, such that all monomials $\mathcal{O}_i[\phi]$ of dimension smaller or equal $d$ can be distinguished by their dependence on upper-left square matrices $\phi|_{n\times n}$ of size $n=n(d)$, i.e. the equation
\begin{equation}\label{equ:inversion-of-couplings}
  \sum_{\{i:[\mathcal{O}_i]\le d\}} g_i \mathcal{O}_i[\phi|_{n\times n}]=\mathcal{F}(\phi_{0000},...,\phi_{nnnn})
\end{equation}
can be inverted for the couplings $g_i(\phi_{0000},...,\phi_{nnnn})$, but this inversion is in general not unique. Once a particular inversion is chosen, we can include monomials $\mathcal{O}_j[\phi]$ with $[\mathcal{O}_j]=d+1$ and find $n(d+1)$ such that the analogue of equation (\ref{equ:inversion-of-couplings}) can be inverted for the couplings $g_j$ while the expressions for the $g_i$ are kept at the particular choice made in the previous step. This procedure can then be iterated to include arbitrarily high dimensional monomials. This procedure is however still ambiguous, because (1) we are free to choose any $n(d)\ge n_o(d)$, and $n(d+1),...$ and (2) the inversion of equation (\ref{equ:inversion-of-couplings}) is in general not unique.

To fix this ambiguity, we can think of  two extremal approaches: (1) We can choose the $n(d),n(d+1),...$ to be the smallest numbers that allow for successive invertibility or (2) we can choose $n(d)$ to be of order $k\sqrt{\theta}$. While the first choice is closely related to the identification of couplings in perturbation theory, the second choice is well adapted to an investigation of the full theory space in the UV-limit, and will be used here. Thus, after manipulating the flow equation to integrate the trace and expand it on the $\mathcal{O}_i $ basis, we will set $n(d)\approx k\sqrt{\theta}$ to match the monomials on the left and right hand sides; this is consistent with the choice of the regulator (\ref{equ:generic-regulator}) and the overall cutoff condition $n<\Lambda\sqrt{\theta}$, $\Lambda\geq k$, and amounts to setting $\Lambda \approx k$.

\subsection{Theory Space and Canonical Dimension}\label{sec:theory-space}

Exact renormalization group equations in general do not preserve a particular action, but rather define a nontrivial flow in a space of functionals of a certain field and symmetry content. One therefore studies paths in this {\emph {theory space}} that are solutions to the flow equation. The properties of the RG-flow may depend on details in  definition of the theory space (compare \cite{machado:2007} for the case of gravity), which is why we will spend some time to examine it in our case.

We need a theory space that is stable under the flow equation and contains the self-dual Grosse-Wulkenhaar action. Such a space was investigated in the matrix base\footnote{Presumably, if one did the same in the position representation, it would be possible to find an isomorphism linking the two constructions, and mapping each RGs flow into each other. However we do not investigate this issue.} \cite{Gurau:2009ni}. For our purposes we can start with the field monomials contained inthe self-dual Grosse-Wulkenhaar action $S$ and investigate which field monomials are generated on the RHS of the flow equation in a vertex expansion. This leads a larger set of field monomials which we insert into the flow equation and the iterate the process until no new field monomials are generated on the RHS of the FRGE. This yields a theory space that is \emph{the smallest ansatz of functionals of matrices $\phi$ that is a priori stable under the RG flow and contains $S$}.

For the self-dual model with a diagonal regulator (\ref{equ:generic-regulator}) we must at least include any even (due to the symmetry $\phi\leftrightarrow-\phi$) polynomial in the field with arbitrary insertions of $K$ operators; a candidate ansatz for the effective action is then
\begin{equation}\label{equ:ansatz-planar}
  \Gamma^{\mathrm{plan}}_k[\phi]=\nu \sum_{i,n_1...n_i} g_{i,n_1...n_{i-1}} \tr\left(\underbrace{\phi.K^{n_1}.\phi\,...\,.\phi.K^{n_i}}_{i\ {\rm terms}}  \right),\ \ \ \ i\ \mathrm{even},
\end{equation}
modulo cyclic permutations of the indices $n_1,...n_i$. We omit the zeros in $g_{i,n_1...n_{i-1}}$ when there is one or no $n_k\neq0$. However, this ansatz excludes operators of the form $\tr\phi^n\,\tr\phi^m$. Such terms are generated, from (\ref{equ:vertex-expansion}) taking $P$ to be scalar\footnote{The matrix $P_{abcd}$ for (\ref{equ:ansatz-planar}) is diagonal, but not scalar $P_{abcd}=\delta_{ad}\delta_{cd}g(a,b)$; however this is irrelevant for the present discussion.}, we get
\begin{equation}\label{equ:vertex-expansion2}
  \partial_t \Gamma_k[\phi]=\frac 1 2 \left(P\delta_{ad}\delta_{bc}-P^2F_{abcd}+P^3F_{abrs}F_{srcd}...\right)\,\dot{R}_k\,\delta^{ad}\delta^{bc}.
\end{equation}
Since $F$ includes the Hessian of the 4-points function $\frac{g_4}{4!}\phi^4$, we have
\begin{equation}\label{equ:hessian-phi4}
F_{abcd}=\nu\frac{g_4}{6}\left(\delta_{cb}\sum_{m=0}^{\Lambda^2\theta}\phi_{dm}\phi_{ma}+\delta_{da}\sum_{m=0}^{\Lambda^2\theta}\phi_{cm}\phi_{mb}+\phi_{da}\phi_{bc}\right)+\dots\;,
\end{equation}
Using only the general form (\ref{equ:generic-regulator}, \ref{equ:ansatz-planar}), we see (modulus numerical constants) that the trace of $F^2$ yields
\begin{equation}
\frac{g_4^2}{k^4}\sum_{a,b}^{\sim k^2\theta}{\textstyle \!\!f\left(\frac{a}{\theta},\frac{b}{\theta}\right)}\!\!\left[\delta_{aa}\!\!\sum_{m,n,k}^{\Lambda^2\theta}\phi_{bm}\phi_{mn}\phi_{nk}\phi_{kb}+\sum_{m,n}^{\Lambda^2\theta}\phi_{am}\phi_{ma}\phi_{bn}\phi_{nb}\right]+\dots\;.
\end{equation}
The first term, evaluated at $\Lambda=k$, yields $\frac{g_4^2}{k^4}\sum_{a}^{k^2\theta}f(\frac{a}{\theta},0)^2 \tr \phi^4$, that is $g_4^2\theta^2 \tr \phi^4$, whereas the latter gives $g_4^2 k^{-4}\tr\phi^2\,\tr\phi^2$; observe that the latter term is suppressed by $\theta^{-2}k^{-4}$ for large $k$. Similar terms will arise from any contraction that does not end up giving a free $\delta$-function, as we can see e.g. by taking the trace of (\ref{equ:hessian-phi4}). These contributions may be understood by a diagrammatic representations in \emph{ribbon graphs}, where the terms with multiple traces are generated from single trace vertices through non-planar diagrams\footnote{In a slight abuse of nomenclature we will call single trace functionals planar and multi trace monomials non-planar as single trace monomials are generated by planar diagrams from single trace vertices while multi trace monomials are generated from single trace vertices by nonplanar diagrams in perturbation theory.}. In fact,  as ordinary commutative theories can be given the familiar representation in Feynman graphs, starting from their momentum representation $\phi_p$, the same can be done here. Thinking of the matrix basis as an analogue of momentum space $\phi_{mn}\leftrightarrow\phi_{p}$ where now each field carries two indices. It is possible to introduce a representation of the perturbative series in graphs with double lines (ribbons), see e.g. \cite{grossewulkenhaar:2005a}. The propagator connects two pairs of indices, whereas the $\phi^4$ vertex has eight free indices. This new structure allows for a richer topology, and most notably for the mentioned non-planar terms. While we will borrow the related vocabulary, we will not need to consider the topological properties of the perturbative expansion.

Let us denote the coupling constants of terms with two traces as $g_{i,\{n\}|j,\{m\}}$, $i+j$ even, and consider a larger ansatz
\begin{align}\label{equ:one-nonplanar-space}
\Gamma'_k&=\Gamma_k^{\mathrm{plan}}\!\!+\nu\theta^{-2}\!\!\!\!\!\sum_{i,j,\{n\},\{m\}}\!\!\!\! g_{i,\{n\}|j,\{m\}} \tr\left(\underbrace{\phi...\,.\phi.K^{n_i}}_{i\ {\rm terms}} \right)\tr\left(\underbrace{\phi...\,.\phi.K^{m_j}}_{j\ {\rm terms}} \right).
\end{align}
 If we do so, the flow will present several new features.  First, the non-planar terms may contribute to the flow of the planar ones. In fact, starting from a $\Gamma_k$ that includes $\nu g_{2|2}\theta^{-2}\tr\phi^2\tr\phi^2=\nu g_{2|2}\theta^{-2}\sum \phi_{mn}\phi_{nm}\phi_{kl}\phi_{lk}$, we get
\begin{equation}
F_{abcd}=\dots+\nu g_{2|2}\theta^{-2}\left(2\delta_{ab}\delta_{cd}\sum_{p,q}^{\Lambda^2\theta}\phi_{kl}\phi_{lk}+ 2\phi_{ba}\phi_{dc}\right).
\end{equation}
Such terms both yield contributions proportional to  $\tr\phi^2$ when traced over. However we have
\begin{equation}
\theta^{-2}\tr[\delta_{ab}\delta_{cd}]\tr\phi^2\approx k^4\tr\phi^2,\ \ \ \ \ \theta^{-2}\tr[\phi_{ba}\phi_{dc}]\approx\theta^{-2}\tr\phi^2,
\end{equation}
so that such traces would contribute to the beta function of the corresponding planar term (in this case, the mass) with a UV suppression of $\theta^{-2}k^{-4}$ and  $\theta^{-4}k^{-8}$ respectively (at $\Lambda=k$). This is a general feature of matrix models.

Furthermore, by considering e.g. $\Gamma_k=...+\nu g_{2|6}\theta^{-2}\tr\phi^2\tr\phi^6$ we also generate terms with three traces, of the form $g_{2|6}k^{-2}\theta^{-2}\tr\phi^2\tr\phi^2\tr\phi^2$. The same reasoning can be repeated to generate terms with an arbitrary number of traces, and it follows that terms with $n+1$ traces appear at order $\nu \theta^{-2n}$. This discussion can be extended to accommodate  for arbitrary insertions of operators $K^n$. In general, several contractions generate terms scaling with non-positive powers of the scale $\theta$.

We conclude that we have to include terms with an arbitrary number of traces, with the only constraint that the \emph{total} number of field operators is even:
\begin{align}\label{equ:full-theory-space}
\Gamma_k&=\Gamma_k^{\mathrm{plan}}\!\!+\nu\theta^{-2}\!\!\!\!\!\sum_{i,i,\{n\},\{m\}}\!\!\!\! g_{i,\{n\}|j,\{m\}} \tr\left(\phi...K^{n_i} \right)\tr\left(\phi...\,\phi.K^{m_j} \right)+\\
\nonumber
&+\nu\theta^{-4}\!\!\!\!\!\!\!\!\!\!\!\sum_{i,i,k,\{n\},\{m\},\{p\}}\!\!\!\!\!\!\!\!\! g_{i,\{n\}|j,\{m\}|k,\{p\}} \tr\left(\phi...K^{n_i} \right)\tr\left(...\right)\tr\left(\phi...\,.\phi.K^{p_k} \right)+\dots\;,
\end{align}
up to cyclic permutations of each string $\{n\},\{m\},...$.  and of arbitrary permutations of the couples $(i,\{n\})$ with $(j,\{m\})$, and so on. By construction (\ref{equ:full-theory-space}) is stable under the RG flow in the vertex expansion, and it is this a theory space that we will use to investigate the self dual GW model.

Let us now consider the canonical dimension of operators, starting with the monomials appearing in $S$. In the position representation (\ref{equ:GW-action-position}) we immediately read off that $\phi(x)$ has mass dimension $1$, the kinetic term $\partial^\mu\partial_\mu+\Omega^2 \tilde{x}^\mu\tilde{x}_\mu$ has dimension $2$ and the volume element $d^4x$ dimension $-4$. As remarked, by construction this translates to $\phi_{mn}$, $K$ and $\nu$ respectively. Since the action must be dimensionless, the dimension of any planar operator, and hence of its coupling is determined by the number of fields $\#\phi$ and of kinetic operators $\#K$ it involves (in agreement with the more rigerous discussion in \cite{Gurau:2009ni}):
\begin{equation}
  \left[g_{i,\{n\}}\right]=-\#\theta\left[\theta\right]-\#\phi\;\left[\phi\right]-\#K\;\left[K\right]=4-i-2\sum_{k=1}^{i-1}n_k.
\end{equation}
For the non-planar terms, we have to take into account the different powers of $\theta$ appearing; for instance $
  \left[g_{i,\{n\}|j,\{m\}}\right]=-j-i-2\sum_{k=1}^{i-1}n_k-2\sum_{h=1}^{j-1}m_k$, and so on. A detailed discussion of the dimension of more general operators can be found in \cite{Gurau:2009ni}.

\subsubsection*{Comment on the Norm on Theory Space}

Notice that we have omitted to define the combinatorial coefficients in front of the couplings $g_{i,\{n\}}, g_{i,\{n\}|j,\{m\}}$ etc. This would make no difference if we had finitely many of those, but in our case their asymptotic behavior in $i,m$ is related to the choice of the norm in the theory space. We expect that not all norms lead to a well-defined flow equation. It is outside the scope of this paper to construct such a norm or even prove existence of a norm in which the FRGE is regular on the respective Banach completion. Instead, we assume that the couplings are asymptotically constant\footnote{It is even possible for our purpose to relax this a bit and allow for some non-constant  asymptotic behavior.} in $n,m$, i.e. that a term such as $\tr[\phi^i.K^n]$ and $\tr[\phi^i.K^{n'}]$ come with the same coefficient\footnote{For convenience, we take $g_{4}\mapsto\frac{1}{4!}g_{4}=\frac{1}{4!}\lambda$ and $g_2\mapsto\frac{1}{2}g_2=m^2$ to make contact with the notation used in perturbation theory.}. Then, we will assume that in this scenario the FRGE can be exists and has a unique, regular flow and proceed to bound the flow to the IR by summing up an infinite number of contributions appearing in the FRGE. Under these assumptions, we will find that the flow can be bounded. In this light, our manipulations are formal, in the same sense as proving a result to all order in perturbation theory is formal if one does not provide a proof of (Borel) summability of the perturbative series.

\subsection{Fixed Point Condition and Asymptotic Safety}\label{sec:fundamental}

Functional renormalization group equations are \textsl{a priori} concerned with effective field theories at a finite renormalization scale and not with fundamental theories at an infinite bare scale. A sufficient condition for a fundamental theory to exist in the exact renormalization group setting is to be able to take the limit in which the renormalization scale goes to infinity. The simplest scenario is to consider points which are left invariant by the RG flow; however, the concept of a \lq\lq fixed point\rq\rq\ is a subtle one in presence of an external dimensionful scale, as argued in \cite{Gurau:2009ni} . It is therefore necessary for our purposes to give a precise definition within the FRGE framework. We start by briefly discussing the floating point condition formulated in \cite{Gurau:2009ni}  in the context of Polchinski's equation; that argument naturally generalizes to any renormalization approach concerned with the Wilsonian action $S_{k,\Lambda}$; we then change the viewpoint slightly and discuss this condition within the FRGE framework.

The Wilsonian effective action is defined so that the partition function
\begin{equation}
  e^{W_k}=Z_k=\int D_\Lambda\phi \exp\left(-S_{k,\Lambda}[\phi]-\Delta_k S[\phi]\right)\;,
\end{equation}
of a field theory remains unchanged when the scale $k$ IR-regulator $\Delta_k S$ is changed. This can be used to derive an exact renormalization group equation for $S_{k,\Lambda}[\phi]$, at a fixed cutoff. In absence of an external scale $L$ one can formulate a sufficient condition for the limit $\Lambda \to \infty$ to exist for $S_\Lambda$, based on dimensional analysis. Let us expand
\begin{equation}
S_{k,\Lambda}[\phi]=\sum_i \hat g^{i,\Lambda}_k k^{\left[\mathcal{O}_i\right]} \mathcal{O}_i[\phi],
\end{equation}
 where $\mathcal{O}_i[\phi]$ are a dense linearly independent set of operators of mass dimension $-\left[\mathcal{O}_i\right]$ in the theory space and $\hat g^{i,\Lambda}_k$ are the associated dimensionless couplings. We can safely take the limit $k\to\Lambda$, getting $S_\Lambda[\phi]=\sum_i \hat g^{i,\Lambda}_\Lambda \Lambda^{\left[\mathcal{O}_i\right]} \mathcal{O}_i[\phi]$. Since the $\hat g^{i,\Lambda}_k$ are dimensionless, $k\partial_k \hat g^{i,\Lambda}_k=0$ implies that $\hat g^{i,\Lambda}_k$ is independent of $\Lambda$ as well. Hence a sufficient condition for $\lim_{\Lambda\to \infty} S_{\Lambda,\Lambda}$ to exist is that $k\partial_k \hat g^{i,\Lambda}_k=0$ for all dimensionless coupling constants\footnote{Strictly, not all beta functions have to vanish, in particular one does not have to require that the wave function renormalization stops running \cite{Rosten:2010vm}.}. The natural generalization of this reasoning to theories with a dimensional external scale is to expand $S_{k,\Lambda}[\phi]=\sum_i \hat g^{i,\Lambda}_k k^{\left[\mathcal{O}_i\right]} \mathcal{O}_i[\hat L,\phi]$, where $\hat L=L k^{-\left[L\right]}$, and impose the floating point condition
\begin{equation}\label{equ:floating-point}
  \left.k \partial_k \hat{g}^{i,\Lambda}_k\right|_{\hat L}=0\,.
\end{equation}
So, although $L$ is the external scale in the theory, one holds $\hat L$ fixed to find a condition to be able to take the limit $\Lambda \to \infty$.

Let us now consider the effective average action $\Gamma_k$. In the FRGE approach, we  can think of $\Gamma_k$ as \emph{defined by the flow equation}. As discussed,  by an appropriate choice of the IR-regulator $\Delta_k S$, the latter is independent of the UV cutoff $\Lambda$ for $k\leq\Lambda$.  Let us expand $\Gamma_k[\phi]=\sum_i \hat g^i_k k^{\left[\mathcal{O}_i\right]} \mathcal{O}_i[\phi]$ in a dense set of linearly independent operators $\mathcal{O}_i$ as above. We use (assume) that only bounded functions of $n$-point functions are observable. Dimensional analysis then implies that an observable $\mathcal{O}$ is a function $\mathcal{O}=k^{\left[\mathcal{O}\right]} f(\hat g_i)$ of the dimensionless couplings $\hat{g}^i$, such that the existence of all dimensionless couplings $\hat g_k^i:=g_k^i k^{\left[\mathcal{O}_i\right]}$ implies that all physical observables exist. If we define the existence of a field theory at a renormalization scale if and only if all dimensionless ratios of observables exist, then a sufficient condition for a fundamental theory is the usual RG-fixed point condition for all essential dimensionless couplings
\begin{equation}\label{equ:fixed-point}
  k \partial_k \hat{g}^{i,\Lambda}_k=0\,.
\end{equation}
We can use this definition even if the system admits a dimensional scale $L$, and take the powers of $L$ occurring in $\mathcal{O}_i$ simply to contribute the dimension of $\mathcal{O}_i$, provided they are well defined when the cutoff is removed. This amounts to showing that it is possible to construct a norm for the $\mathcal{O}_i$ that is well defined even when the cutoff is removed. In the Grosse-Wulkenhaar model, where the external scale is $\theta$, we can set $\hat{\theta}=\theta\Lambda^2$, and we must require that the norm of the operators does not diverge when the cutoff is removed, i.e. $\hat{\theta}\to\infty$. This is indeed the case  in the GW model with our theory space because, as we have remarked, the flow generates only contributions that scale with with non-positive powers of $\hat{\theta}$ in a vertex expansion; for a more detailed discussion of this point, see \cite{Gurau:2009ni}.  With respect to the condition (\ref{equ:fixed-point}) one sees that the free theory, defined by the massless quadratic part of (\ref{equ:GW-action-matrix}), is indeed a fixed point of the RG flow.

Practically there is no difference between the floating point condition (\ref{equ:floating-point}) and the fixed point condition (\ref{equ:fixed-point}) on our theory space. The different reasoning comes from the different ways in which we implement the requirement to be able to safely take the renormalization scale to infinity.

Once we have identified a fixed point, we can study the behavior of the couplings in its neighborhood. This is particularly simple at the Gaussian fixed point (GFP), the one corresponding to the free massless theory. Let us limit our analysis to the vicinity of the GFP, for the sake of the discussion. We can distinguish between the couplings that have positive mass dimension (relevant couplings), the ones with negative mass dimension (irrelevant), and the dimensionless ones (marginal); in the self dual GW model there is just one essential marginal coupling (i.e. after absorbing of the wave function normalization in a field redefinition), that is $\lambda$. If we give an initial condition in the UV for the theory, and integrate back to the IR to find the full effective action, we cannot include any irrelevant coupling, in absence of a non-Gaussian fixed point. On the other hand, marginal and relevant couplings will have a flow, and also generate terms corresponding to the irrelevant couplings as effective vertices.

A subtle point of this local analysis is to understand what happens to the flow of the marginal coupling. When we consider quantum corrections, it may happen that it behaves as a relevant coupling, flowing away from the GFP in the IR (asymptotic freedom), or as an irrelevant one, which makes it impossible to have a non-trivial interacting theory. This seems to happen in the commutative $\phi^4$ theory. Finally, it may remain marginal, i.e. have just a small finite flow that does not drive it to zero in the IR. This is what is known to happen in perturbation theory for the GW model, and implies that the running of $\lambda$ as a function of the energy scale is bounded in a neighborhood of the GFP. We refer to this scenario as \emph{asymptotic safety} (compare \cite{disertori:2007}).

\section{Beta functions}\label{sec:beta-function}

We present here the computation for the $\beta$-function for the coupling of the $\phi^4$ interaction ($\lambda$ or $g_4$). This will also require considering the field strength renormalization $\eta$. We first perform a simple calculation for $\Gamma_k:=S$, finding that in this truncation $g_4$ is constant, i.e. the beta function vanishes. Since the FRGE leads to an infinite set of coupled $\beta$-functions, we extend the calculation we discuss which terms can \textsl{a priori} appear in the $\beta$-function for $g_4$ and in $\eta$. We then bound such contributions to find that the flow of $\lambda$ is finite, i.e. the theory is \emph{asymptotically safe}.

\subsection{Beta functions in the truncated self-dual model}\label{sec:oneloop-dual}

As a preliminary computation, we take an ansatz consisting only of the power-counting relevant and marginal couplings(GW truncation), i.e.
\begin{equation}\label{equ:gamma1loop}
\Gamma_k[\phi]=\nu\left(\frac{1}{2}\mathcal{Z}\phi.K.\phi+\frac{g_2}{2}\mathcal{Z} \phi.\phi+\frac{g_4}{4!}\mathcal{Z}^2\phi.\phi.\phi.\phi \right),
\end{equation}
where $\mathcal{Z}$ is the field-strength renormalization, and all the couplings depend on the scale $k$: $\mathcal{Z}\equiv\mathcal{Z}(k)$, $g_{n}=g_n(k)$; where as above $\eta:=d_t\log\mathcal{Z}$. Using this ansatz in (\ref{equ:vertex-expansion}) yields the saddle-point approximation of the IR modified path integral. This is sometimes referred to as a \lq\lq RG improved\rq\rq\ one-loop computation. However, one should bear in mind that this treatment genuinely differs from its perturbative counterpart, although, as we are about to see, this yields the same results at first-order.

Let us use the regulator (\ref{equ:regulator-sharp}), so that $G(a,b)=[g_2+\frac{4}{\theta}(a+b+2) +\alpha k^2] H(k^2\theta-\beta\|a,b\|)$. The expansion of the trace yields, in the planar sector and up to $\phi^4$,
\begin{equation}\label{equ:trace-w-PR}
\dot{\Gamma}_k=\!-\!\sum_{a,b}^{\infty}\!\!\frac{g_4\dot{R}(a,b)}{6\nu G(a,b)^2}\,2\delta_{aa}(\phi^2)_{bb}\!+\!\sum_{a,b}^{\infty}\!\!\frac{g_4^2\mathcal{Z} \,\dot{R}(a,b)}{36\nu G(a,b)^2}\,2\delta_{aa}\sum_{c}^{\Lambda^2\theta}\!\!\frac{(\phi^2)_{bc}(\phi^2)_{cb}}{G(a,c)},
\end{equation}
where the factors of two account for the symmetry $a\leftrightarrow b$. Observe that the $\delta$-functions coming from $\dot{R}$ can either contribute to boundary terms (when acting on the index $b$), or yield terms suppressed by $\theta^{-2} k^{-4}$ (when acting on $a$). We do not need to consider the former case because these lie outside our truncation. For the latter, the one-loop vanishing of the beta-function will straightforwardly hold also for them. Let us evaluate the flow equation at $\Lambda=k$, and  indicate the terms outside the truncation with dots:
\begin{align}\nonumber
\dot{\Gamma}_k=&-\sum_{a,b}^{\beta k^2\theta}\frac{g_4\mathcal{Z}\alpha(1+\frac{\eta}{2}) k^2}{6G(a,b)^2}\,2\delta_{aa}(\phi^2)_{bb}\;+\\
\label{equ:trace-w-P}
&+\sum_{a,b,c}^{\beta k^2\theta} \frac{g_4^2\mathcal{Z}^2\alpha(1+\frac{\eta}{2}) k^2}{36G(a,b)^2}\,2\delta_{aa}\frac{(\phi^2)_{bc}(\phi^2)_{cb}}{G(a,c)}+...\;.
\end{align}
Since $G$ is diagonal, but not scalar, we have a residual dependence on the internal index $c$. To extract the term proportional to $\tr\phi^4$ we have to consider the terms where the fields $\phi$ are not contracted with any index-dependent function. This amounts to restricting the sums to $G(a,b)|_{b=0}$. We obtain
\begin{equation}\label{equ:1loop-vanishing1}
\frac{\nu}{4!}\left(2\eta \hat{g}_4 +d_t \hat{g}_4\right) \tr\phi^4=(1+\frac{\eta}{2})\frac{\hat{g}_4^2}{18}\sum_{a_1,a_2}^{\beta k^2\theta}\frac{\delta_{a_1a_1}\delta_{a_2a_2}}{G(a,0)^3}\tr\phi^4\;,
\end{equation}
where the hat denotes dimensionless variables. The contribution of the term containing $\delta$-functions is obtained from the above by the sum over $a_1,a_2$ with the evaluation at $\beta k^2\theta$.

Similarly, we get the running of the mass squared $g_2$:
\begin{equation}
\frac{\nu}{2}(\eta{g}_2+2\hat{g}_2+d_t\hat{g}_2)\tr\phi^2=-\alpha(1+\frac{\eta}{2})\frac{\hat{g}_4}{3}\sum_{a}^{\beta k^2\theta}\frac{\delta_{aa}}{G(a,0)^2}\tr\phi^2,
\end{equation}
where the  contribution $2\hat{g}_2$ to the l.h.s. comes from the canonical dimension of the coupling constant: $g_2=k^2\hat{g}_2$.

As for $\eta$,  the kinetic term has the form $\tr[\frac{4}{\theta}(a+b+2)\phi_{ab}\phi_{ba}]=\tr[\frac{8}{\theta}(b+1)\phi^2_{bb}]$, and defining $\bar{b}:=\frac{8}{\theta}(b_1+b_2+1)$,
\begin{align}\nonumber
\nu\frac{\eta}{2}\tr[\bar{b}\,\phi^2_{bb}]&=-(1+\frac{\eta}{2})\frac{g_4\alpha}{3}\sum_{a,b}^{\beta k^2\theta}\frac{\partial}{\partial \bar{b}}\!\!\left.\frac{\delta_{a_1a_1}\delta_{a_2a_2}}{G(a,b)^2}\right|_{b=0}\tr[\bar{b}\,\phi^2_{bb}]=\\
\label{equ:1loop-vanishing2}
&=(1+\frac{\eta}{2})\frac{g_4\alpha}{3}\sum_{a}^{\beta k^2\theta}\frac{\delta_{a_1a_1}\delta_{a_2a_2}}{G(a,0)^3}\; \tr[\bar{b}\,\phi^2_{bb}]\;.
\end{align}
Again, the contribution of the non-extensive terms can be obtained by replacing the sum with the evaluation. Putting the two equations for $\eta$ and $d_t\hat{g}_4$ together, it follows that
\begin{equation}\label{eq:vanishingoneloop}
d_t\hat{g}_4=0\ \ \ \mathrm{in\ the \ GW \ truncation},
\end{equation}
and we have recovered the \emph{one-loop vanishing of the $\beta$-function}. Furthermore,
\begin{equation}
\eta=\frac{\hat{g}_4}{6\pi^2}\frac{\alpha\beta^2(1+\frac{1}{2}\eta)}{\textstyle 2(\alpha+\hat{g}_2)(\alpha+\hat{g}_2+4\beta)^2}+O(\frac{1}{k^2\theta})\approx\frac{\hat{g}_4}{12\pi^2}\frac{\beta^2}{\textstyle (\alpha+4\beta)^2}+O(\frac{1}{k^2\theta}).
\end{equation}
The terms $O(\frac{1}{k^2\theta})$ arise from replacing the sums with integrals and from the $\delta$-functions. We develop the result to the zeroth order in $\hat{g}_2=g_2/k^2\ll1$, and we neglect the extra $\eta$-dependence coming from the regulator, which amounts to linearizing the result in $\hat{g}_4$. The result depends on the choice of the regulator, because the cutoff is not \lq\lq optimal\rq\rq; however, in the limit $\alpha\to0$ the IR regulator is removed, and we find $\eta=\hat{g}_4/192\pi^2$, in agreement with \cite{grossewulkenhaar:2004,Gurau:2009ni}\footnote{One must take into account the different conventions for the scale derivative and $\eta$ in the latter reference.}. As a further check, we can compute $d_t\hat{g}_2$, expanding it up to the first order in $\hat{g}_2$. In the limit $\alpha\to0$ we are left with the universal logarithmic divergence, yielding
\begin{equation}
d_t\hat{g}_2\xrightarrow{\alpha\to0}-2\hat{g}_2-\eta\hat{g}_2+\frac{ \hat{g}_4\,\hat{g}_2}{96\pi^2}+O(\frac{1}{\theta k^2})=-2\hat{g}_2+\frac{ \hat{g}_4\,\hat{g}_2}{192\pi^2}+O(\frac{1}{\theta k^2}).
\end{equation}
This agrees with the logarithmic part of \cite{grossewulkenhaar:2004} up to a sign which we believe to be a typo in the original manuscript. Remark that even if the explicit result is scheme dependent, the vanishing of the $\beta$-function in this truncation is not, as in (\ref{equ:1loop-vanishing1}--\ref{equ:1loop-vanishing2}) $\alpha,\beta$ and $\hat{g}_2$ are generic; furthermore, the result holds at all orders in $\theta k^2$.

Let us repeat the same computation with the \lq\lq optimized\rq\rq\ regulator (\ref{equ:regulator-litim}), denoting the new result by a prime. One sees that in the approximation (\ref{equ:gamma1loop}) the trace cannot yield any index-dependent term other than the ones coming from $\dot{R}$ (\ref{equ:rdot-litim}) (proportional to $\eta'$), so that
\begin{equation}\label{equ:eta-litimR}
\nu\frac{\eta'}{2}\tr[\bar{b}(\phi^2)_{bb}]=\eta'\;\sum_{a,b}^{\beta \theta k^2}\frac{\hat{g}'_4}{6(\beta+\hat{g}'_2)^2} \delta_{aa}(\phi^2)_{bb}\frac{\bar{b}}{8}=\frac{\eta'\ \beta^2\;\hat{g}'_4}{96(\beta+\hat{g}'_2)^2}\tr[\bar{b}(\phi^2)_{bb}].
\end{equation}
Therefore, for generic values of $\hat{g}'_n$, the above equation is satisfied only for $\eta'=0$, apparently contradicting the previous result. Then,
\begin{align}\nonumber
d_t\hat{g}_4'&=\frac{(\hat{g}_4')^2}{3\pi^2}\!\!\sum_{a_1,a_2}\!\!\!\frac{\beta}{(\hat{g}_2'+\beta)^3}H\left({\textstyle \beta k^2-\frac{4}{\theta}(a+1)}\right)+O(\frac{1}{k^2\theta})\approx\\
\label{equ:beta-litimR}
&\approx \frac{(\hat{g}_4')^2}{96\pi^2}+O(\frac{1}{k^2\theta}),
\end{align}
where we have not written any contribution in $\eta'=0$. This result is already independent of the scheme parameter $\beta$. For the mass, we get a $\beta$-dependent quadratic divergence and the logarithmic divergence
\begin{equation}
d_t\hat{g}_2'=-2\hat{g}_2' -\frac{\beta\hat{g}'_4}{192\pi^2}+\frac{\hat{g}_2' \;\hat{g}_4' }{96\pi^2}+O(\frac{1}{k^2\theta}).
\end{equation}
This results shows that the two sets of $\beta$-functions differ by a redefinition of the field, $\mathcal{Z}^{1/2}\phi\to\phi'$, where $\mathcal{Z}$ depends on $k$. By performing such a transformation, one recovers the previous result. Even if the latter choice for the regulator does not relate immediately to perturbation theory, we will use it in the following as it reduces the dependence on scheme parameters and offers simpler computations.

\subsection{Full Vertex Expansion}\label{sec:full-vertex-exp}

We now examine the vanishing of the beta function for $g_4$ to all orders of perturbation theory with FRGE tools. For this we have to consider the full theory space (\ref{equ:full-theory-space}). However, within the FRGE framework we won't exactly have that $d_t \hat{g}_4=0$, but rather find that the coupling will have a bounded RG flow. This will require to bound the scaling of both of the power-counting marginal couplings $d_t\hat{g}_4$ and $\eta$, as we evolve them from the UV to the IR. For this purpose, it is useful to introduce $\tau=-t$, the RG time towards the IR. Our manipulations are rather formal, in the sense that they rely on the assumption of existence and regularity of the flow in the whole theory space. 

\subsubsection*{Preparations}

We consider the sector of the theory space given by the irrelevant terms and consider its linearized flow, regarding the relevant and marginal couplings as independent parameters. We suppose that the space of the irrelevant couplings (which is a linear subspace of finite codimension of the theory space) can be given a Banach space structure, and that the flow defined by the above equations is regular and differentiable in the Banach norm. Furthermore, we suppose that the dependence of the flow on the relevant and marginal couplings, treated as external parameters, is differentiable as well. The scaling of an irrelevant coupling  at the free-field fixed point, when we also set the marginal $\phi^4$ coupling to zero, will be dominated by its mass dimension $[g_{n}]<0$:
\begin{equation}
d_\tau\hat{g}_{i,n}=[g_{i,n}]\hat{g}_{i,n}+\left(\mathrm{terms\ from\ order\ of\ }\phi^{i+2}\right)+\mathrm{h.o.\ terms},
\end{equation}
where the terms from order $i+2$ are proportional to $\hat{g}_{i+2, n}$, and to similar nonplanar terms, with a combinatorial coefficient that depends on $i$ but not on $n$. Their presence forbids to straightforwardly diagonalize the linear flow on the base we have chosen. However, since all these terms have a smaller (negative) mass dimension than $\hat{g}_{i,n}$, they cannot spoil the exponential suppression\footnote{Strictly speaking, this is an assumption depending on the Banach space topology. The assertion is true for finite dimensional spaces; here it can not be proven straightforwardly because the $g_{i+2,n}$ in turn depend on an infinite number of couplings.} given by the ratio $[g_{i,n}]$.  Using the stable manifold theorem, (e.g. Theorem 6.1 in \cite{Rouelle:1989}) the flow of the linearized system is in a neighborhood of the fixed point (where also the external parameters are set to vanish) conjugated to the flow of the full system by a diffeomorphism, such that
\begin{equation}\label{equ:bound-irrelevant}
\left|\hat{g}_{n}(\tau)\right|\leq \kappa_n \exp\left[\tau[g_{n}](1-\varepsilon/2)\right],\ \ \ \ \tau>0,
\end{equation}
for some $\kappa_n>0$ and any $\varepsilon>0$. Since the flow depends smoothly on the relevant couplings, the same estimate will hold in neighborhood of the fixed point in the full theory space, possibly shrinking the neighborhood for the irrelevant couplings. Furthermore, when $\varepsilon$ is small enough such a neighborhood is stable under the RG flow towards the IR, at fixed $\hat{g}_2$ and $\hat{g}_4$. For simplicity, we can take $\kappa_n=1$. Then, calling $r$ the radius of the neighborhood where the above analysis valid, we can consider the estimate to hold for $\tau>\tau^*:=|\log r|$, so that the smaller the neighborhood is, the larger $\tau^*$ we will have to be taken.

Let us summarize the picture. Arguing by the stable manifold theorem, we know that at $\hat g_4=0$ there is some one-dimensional critical manifold, that  is diffeomorphic to what we find in the linearized case (the half-line $\hat g_0>0$). By regularity in $\hat g_4$, we expect this to hold perturbing a bit away from zero. Since the marginal coupling itself flows, it is not clear that we can consistently set it to be nonzero. In fact, its IR flow may blow up, contradicting our argument, or drive it to zero, yielding a trivial theory. We will use the estimate (\ref{equ:bound-irrelevant}) to show that $\hat g_4$ has a bounded flow. Notice that the estimate relies only on the mass scaling of irrelevant couplings in the IR, so that it holds for any small enough initial condition. However, this does not mean that for any small value of the the couplings we will end up with an asymptotically safe theory! On the contrary, it means if we are sufficiently close to the GFP, regardless of the precise value of the irrelevant couplings, those will only generate a small flow for $\hat g_4$, so that indeed our picture is consistent and even in the full system we  have a one dimensional critical surface emanating from the free theory. This will be a deformed image of what found at $\hat g_4=0$ in the linearized case, and its precise shape will depend on the regularization scheme employed\footnote{In this light, the perturbation theory proof of \cite{disertori:2007} employed the scheme that yields a critical surface exactly on the $\hat{g}_4>0$ segment, which still differs from the linearized case where we had the whole half-line.}.  Only the theories defined on this critical surface will be asymptotically safe.

\subsubsection*{Bound Estimate}

Let us now consider the beta functions for the power-counting marginal operators. We will outline the general idea how these can be bounded here and refer to the appendix for more details.

The vertex expansion combined with the one-loop structure of the FRGE implies that not all irrelevant couplings constants can appear in the beta functions of $d_\tau \hat{g}_4$ and $\eta$. In particular, we can limit ourselves to consider only those couplings that correspond to operators involving at most four (resp. six) fields in the beta functions of $\mathcal Z$ (resp. $\hat g_4$). However, those can still contain an infinite number of $K$s, so we are still left with an infinite number of couplings, and arbitrarily many products of them. For instance the ones appearing as coefficients of $\tr\phi^4$ are $g_{2,n}$, $g_{4,n}$, $g_{6,n}$ and the non planar ones $g_{1,n|1}$, $g_{1|3}$, $g_{2|2}$, $g_{1|5}$, $g_{2,n|4}$. A detailed discussion of this, explaining why we get less terms that one may na\"ively expect, can be found in the appendix, where the conditions under which the non-planar terms may occur are also discussed. In this scenario it is not clear that all these terms do not add up to give a finite effect. This is what we have to show now.

The r.h.s. of the flow equation splits into a finite number of traces for each coupling,  e.g.
\begin{equation}\label{equ:generic-sum}
\sum_{a_1,a_2}\frac{C\ \dot{R}(a,0)\sum_{p,q=0}^{\infty}\left(\frac{a+1}{\theta}\right)^{p+q}\;C_p\hat{g}_{4,p}\;C_q\hat{g}_{4,q}}{\scriptstyle \left(1+\hat{g}_2+\sum_{r=2}^\infty \left(\frac{a+1}{\theta}\right)^r\hat{g}_{2,r}+g_{1|1}\right)^2\left(1+\hat{g}_2+\sum_{r=2}^\infty \left(\frac{a+1}{\theta}\right)^r\hat{g}_{2,r}\right)}\;\tr\phi^4,
\end{equation}
that generalize the one-loop computation (\ref{equ:1loop-vanishing1}).

Some comments are in order: the coefficients $C_p$ and $C_q$ come from the combinatorics in computing the Hessian for $g_g\tr\phi^4$ and $g_{4,p}\tr[K^p\phi^4]$. Indeed $C_0/C_p=4$ $\forall p >0$, as observed in the Appendix \ref{app:traces}. The structure of the denominator is due to the fact that not any contraction with non-planar terms generate planar ones, see (\ref{equ:planar-projection}) and the subsequent discussion. Similar expressions hold for terms involving $\lambda_{6,n}$, the non-planar couplings and for the ones contributing to $\tr[K\phi^2]$, i.e. to $\eta$. We want to use (\ref{equ:bound-irrelevant}) to perform all these sums, ultimately yielding a non-autonomous differential equation $d_\tau\hat{g}_4=f(\tau)$. The beta functions are composed of a finite number of sums which all have the same structure. We denote a generic sum of the form (\ref{equ:generic-sum}) by $\Xi$, and proceed in several steps.

\par
{\bf Step 1.} We consider $|\Xi|$, and overestimate any combinatorial coefficient $C_p$  with $C_0$; we get rid of such coefficients by including them in $C$. We expand the denominator in powers of $\frac{a+1}{\theta}$,  and reorder the sum as $\sum_{a}\sum_p (\frac{a+1}{\theta})^p F_p(g)$, where $F_p(g)$ indicates the dependence on all the couplings. We take the modulus of each summand (thus overestimating the sum) and swap the summations in $a$ and $p$. We perform the summation in $a=(a_1,a_2)$, which by (\ref {equ:bound-integral}) gives $\sum_p \sum_{a}(\frac{a+1}{\theta})^p F_p(g)\leq\sum_p M_0^p F_p(g)$ for some $M_0>0$; we can also suppose $M_0>1$.

\par
{\bf Step 2.} The function $F_p(g)$ arising from the expansion in powers of $\frac{a+1}{\theta}$ involves powers of the $a$-independent part of the denominator, $\frac{1}{(1+\hat{g}_2+\hat{g}_{1|1})^k}$, with $k\lesssim p$. We can overestimate this by some $M_1^p$. In fact, the mass is positive, and the irrelevant coupling $|\hat{g}_{1|1}|$ is by (\ref{equ:bound-irrelevant}) bounded. Then $\sum_p M^p F_p(g)\leq \sum_p (MM_1)^p \tilde{F}_p(g)$, where $\tilde{F}_p(g)$ depends only on the (planar and non-planar) irrelevant couplings involving positive powers of operators $K$. We now set $M_2:=M_0M_1$.

\par
{\bf Step 3.} The functions $\tilde{F}_p(g)$ contain products of couplings  arranged in such a way that they carry an overall factor of $(\frac{a+1}{\theta})^p$. We want to rearrange the sum in terms of functions $G_{2q}(g)$ such that $q$ labels the combined mass dimension of the couplings. For instance, the product $g_{4,m}g_{4,k}$ would appear in $\tilde{F}_{m+k}$ and has total dimension $-2(n+k)=-2q$, but $g_{2,n}g_{4,m}g_{4,k}$ would appear in $\tilde{F}_{n+m+k}$ with dimension $-2(n-1+m+k)=-2q'$. The two orderings are thus not equivalent. However, a term with mass dimension $-2q$ can give at most a factor of $(\frac{a+1}{\theta})^{2q}$ because the lowest dimension couplings involving powers of $\frac{a+1}{\theta}$ are the $g_{2,2}$. Hence $\sum_p M_2^p \tilde{F}_p(g)\leq\sum_{q} M_2^{2q} G_{2q}(g)$. In what follows it will be important to observe that all the couplings appearing in the sum have a maximum dimension $-d_o$, that is the dimension of the least irrelevant coupling appearing in the sum. For instance, terms proportional to $g_{6,n}$ have $-d_o=[g_6]=-2$, whereas terms proportional to $g_{4,m}g_{4,n}$ have $d_o=0$.

\par
{\bf Step 4.} We use (\ref{equ:bound-irrelevant}) in the inequality, and replace all the irrelevant couplings by their scaling, but  do not touch the marginal couplings $\hat{g}_4$. We factor an $e^{-\tau(d_o-\varepsilon)}$ common to all the summands, obtaining
\begin{equation}
|\Xi|\leq f(\hat{g}_4)\ N_0 e^{-\tau(d_o-\varepsilon)}\sum_{q=0}^\infty e^{-\tau(2q-q\varepsilon)+\mu_0 q}G_{2q}\;,
\end{equation}
where the scaling due to the common mass dimension has also been factored out. $G_{2q}$ is a combinatorial coefficient and $f(\hat{g}_4)$ is quadratic, linear or constant depending on the contribution we are considering. $G_{2q}$ counts in how many ways, allowing for repetitions, one can chose irrelevant couplings that have total dimension $-2q$. This is bounded by the number of ways of distributing $q$ objects in $k(q)$ boxes, where $k(q)$ is the number of distinct couplings that have mass dimension larger than $-2q$. Even when we allow for the mentioned non-planar contributions, this number is still linear in $q$, so that $G_{2q}$ is bounded by a binomial coefficient, and thus by an exponential. Thus we can perform the sum
\begin{equation}
|\Xi|\leq f(\hat{g}_4)\ N_0 e^{-\tau(2q_o-\varepsilon)}\sum_{q=0}^\infty e^{-\tau(2q-q\varepsilon)+\mu q}\lesssim N f(\hat{g}_4) e^{-\tau(2q_o-\varepsilon)},\ \ \ \tau\ \mathrm{large}.
\end{equation}

\par
It is straightforward to check that these arguments go trough for any of the traces appearing in either the coefficient of $\tr\phi^4$ and of $\tr[K\phi^2]$. An explicit calculation is illustrated in full detail in the appendix for one particular contribution.

\par
The behavior of this result depends on whether $d_o>0$. Any term that does not depend quadratically on $\hat{g}_4$ has trivially $d_o\geq2$, because it is expressed by a sum whose highest dimensional couplings are strictly irrelevant. As for the ones quadratic in $\hat{g}_4$, we can separately consider the case involving strictly irrelevant two-point couplings, e.g. $g_{2,n}$ with $n\geq2$, for which again $d_o\geq2$, and the remaining ones, for which we have proven in Section \ref{sec:oneloop-dual} that any contribution to $d_t\hat{g}$ is canceled by a similar one in $\eta$. One finally finds the bound for the $\tau$-dependent differential equation, in terms of the numerical coefficients $c_0, c_1$ and $c_2$:
\begin{equation}
|d_\tau \hat{g}_4|\lesssim e^{-2\tau+\varepsilon \tau}\left(c_0+c_1\hat{g}_4+c_2\hat{g}_4^2\right)\;.
\end{equation}
A differential equation of this kind can be integrated to yield a finite flow for $\hat{g}_4(\tau)$ and, when $\hat{g}_4$ is small enough,  yields an exponentially bounded flow. We have to set the initial condition $\hat{g}_4(\tau^*)$ in the neighborhood where our analysis is valid, i.e. for $\tau^*$ sufficiently larger than zero, and so that  $\hat{g}_4(\tau)$ never leaves that neighborhood. At the same time all irrelevant couplings have to remain on the critical surface. Then (\ref{equ:bound-irrelevant})  bounds the deviation of the critical surface from the one we would have when setting all power-counting irrelevant couplings to zero. Then we have a fully consistent picture. This means that there exists an interval $[0,\epsilon[$ for $\hat{g}_4$  where all the RG effects, including the ones accounted for in higher order effective vertices, just add up to give a finite flow for the coupling. We conclude that in this interval the self-dual Grosse-Wulkenhaar model is asymptotically safe and has a one-dimensional UV-attractor.

Let us comment on the strategy we have adopted, which consists of two ingredients: first, the vanishing the $\beta$-function \lq\lq at one-loop\rq\rq, or rather in a truncation where we consider only the relevant and marginal couplings. Second, the fact that the total contribution of the irrelevant terms is itself irrelevant, in the sense that it can be summed up to give something that has the overall scaling of an irrelevant coupling, i.e a negative exponential. This is not trivial, and requires the number of irrelevant couplings contributing to the running of $\hat{g}_4$ not to grow too fast with their mass dimension; in particular had we found $G_q\approx q!$, all our reasoning would have fallen. Finally, we stress again that to turn this argument in a proper proof, we would have to define a Banach topology of the theory space and proving existence and regularity of the flow, a problem which is likely to be as hard as proving the Borel-summability of the perturbative series, and would require to properly account for the renormalon problem.

\subsection{A Calculation in the Duality-Covariant Model}

The aim of this section is to show how conveniently the FRGE framework can be applied to study more general matrix models. In particular, we will consider the Grosse-Wulkenhaar model without Langmann-Szabo symmetry, i.e. $\Omega\neq1$ in (\ref{equ:GW-action-position}); this allows for a much vaster theory space. However, the one-loop computations, i.e. the ones for which we take the simple ansatz $\Gamma_k:=S$, are easily feasible, and we will indeed reproduce with little effort the lowest order of the $\beta$-functions for the duality covariant model originally derived by Grosse and Wulkenhaar \cite{grossewulkenhaar:2004}.

We want to investigate the matrix model defined by (\ref{equ:GW-action-matrix}) when $\omega\neq0$. Let us rewrite it in terms of the scale dependent couplings
\begin{equation}
  S[\phi_{mn}]=\nu\mathcal{Z}\, \tr\left(\frac{\rho+\frac \omega 2}{2} \phi.\tilde X_\mu.\tilde X^\mu.\phi+\frac \omega 2 \phi.\tilde X_\mu.\phi.\tilde X^\mu +\frac{g_2}2 \phi^2 + \mathcal{Z}\frac{g_4}{4!}\phi^4\right),
\end{equation}
where $\mathcal{Z}\equiv\mathcal{Z}(k)$, $\rho\equiv\rho(k)$, $\omega\equiv\omega(k)$ and $g_n\equiv g_n(k)$. Notice that $\omega$ plays the role of the coupling for the non-diagonal, duality-covariant terms, $\rho$ accounts for the running of the diagonal kinetic term, and $\mathcal{Z}$ is the wave function renormalization. The five couplings correspond to just four independent physical quantities, so that one of them is redundant.

The action now contains the non-diagonal quadratic term
\begin{align}\nonumber
\phi.\tilde{X}_\mu.\phi.\tilde{X}^\mu&=\phi_{mn}\phi_{kl} \mathcal{H}_{mnkl}\;.\\
\label{equ:definition-H}
\mathcal{H}_{mnkl}&=\frac{2}{\theta}\left[\left(\sqrt{(m_1+1)(n_1+1)}\delta_{m_1+1,l}\delta_{n_1+1,k}\;+\right.\right.\\
\nonumber
&\ \ \ \ \ \ \ \left.\left.+\sqrt{m_1\phantom{|}n_1}\delta_{m_1-1,l_1}\delta_{n_1-1,k}\right)\delta_{m_2,l_2}\delta_{n_2,k_2}+\{1\}\leftrightarrow\{2\}\right]\;.
\end{align}
If we were to construct a theory space including such a term, we would have to allow for arbitrary \lq\lq jumps\rq\rq\ in the indices of the form $\delta_{m_1+N,l}\delta_{n_1+N,k}$. However, let us restrict ourselves to the ansatz $\Gamma_k=S_k$, and consider the flow equation (\ref{equ:vertex-expansion}). To avoid the complications of inverting a non diagonal matrix $G=\Gamma^{(2)}|_{\phi=0}+R$, where
\begin{equation}
\left.\Gamma^{(2)}_{mnkl}\right|_{\phi=0}\!\!\!\!=\mathcal{Z}\nu\left[g_2\delta_{ml}\delta_{nk}+\left(1+\frac{\omega}{2}\right)\mathcal{K}_{mnkl}+\omega \mathcal{H}_{mnkl}\right]\;,
\end{equation}
it is sufficient to chose the regulator to be
\begin{equation}\label{equ:regulator-covariant}
R=\nu \mathcal{Z}\left[k^2-\left(1+\frac{\omega}{2}\right)\mathcal{K}-\omega \mathcal{H}\right]\,H\!\left[{\textstyle  k^2-\frac{m+1}{\theta/4}}\right]H\!\left[{\textstyle  k^2-\frac{n+1}{\theta/4}}\right]\;.
\end{equation}
We pay the simple form of $G$ with a complicated dependence of $\dot{R}$ on the $\beta$-functions of the two-point terms:
\begin{equation}
\dot{R}=\eta R\! +\!\nu\mathcal{Z}\left[2k^2-d_t\rho\,\mathcal{K}-d_t\omega \left({\textstyle \frac{1}{2}}\mathcal{K}+\!\mathcal{H}\right)\right]\,H\!\!\left[{\textstyle k^2-\frac{m+1}{\theta/4}}\right]H\!\!\left[{\textstyle k^2-\frac{n+1}{\theta/4}}\right],
\end{equation}
up to contributions proportional to $\delta$-functions, which are not visible in our truncation for analogous reasons as in the self-dual model. All the following calculations are understood to hold up to $O(1/\theta^2k^4)$. We could have chosen the regulator (\ref{equ:regulator-sharp}), that would have given the same manipulations as in \cite{grossewulkenhaar:2004} up to an IR correction of the propagators, linear in $\alpha$.

It is rather straightforward to extract the $\beta$-functions. Observing that $G$ is scalar when the indices are smaller than the UV cutoff, we have that the same holds for $P=G^{-1}$ so that (\ref{equ:vertex-expansion}) reads
\begin{equation}\label{equ:vertex-expansion2}
  \dot{\Gamma}_k[\phi]=\frac 1 2 \left(P\delta_{au}\delta_{bt}-P^2F_{abtu}+P^3F_{abrs}F_{srtu}...\right)\,(\dot{R}_k)_{utcd}\,\delta^{ad}\delta^{bc}.
\end{equation}

The new term in the ansatz is $\tr\left[\phi \mathcal{H}\phi\right]$; however, a direct computation (see the Appendix \ref{app:example}) shows that no such term can be generated from $\tr\left[F\dot{R}\right]$; this can be rephrased in a diagrammatic setting saying that there are no planar tadpoles with a \lq\lq jump\rq\rq. As a result, we get
\begin{equation}\label{equ:omega-zeta}
d_t\left(\mathcal{Z}\frac{\omega}{2}\right) \tr\left[\phi \mathcal{H}\phi\right]=0\ \ \ \ \Rightarrow\ \ \ d_t\omega+\omega\eta=0\;,
\end{equation}
which holds to any order in $\theta k^2$ in this truncation. The traces proportional to $\tr\left[\phi \mathcal{K}\phi\right]=\tr[(\phi^2)_{bb}\bar{b}]$ are generated as in (\ref{equ:eta-litimR}) from the terms in $\dot{R}$ containing $\mathcal{K}$, yielding
\begin{equation}
\frac{\nu}{2}d_t\left(\mathcal{Z}(\rho+\frac{\omega}{2})\right) \tr[(\phi^2)_{bb}\bar{b}]=\left[\eta(\rho+\frac{\omega}{2})+d_t\rho+d_t\frac{\omega}{2}\right]\frac{\mathcal{Z}\ \beta^2\;\hat{g}_4}{96(\beta+\hat{g}_2)^2}\;,
\end{equation}
which can be simplified using (\ref{equ:omega-zeta}):
\begin{equation}\label{equ:rho-zeta}
d_t\left(\mathcal{Z}\rho\right) \tr[(\phi^2)_{bb}\bar{b}]=\left[\eta\rho+d_t\rho\right]\frac{\mathcal{Z}\ \beta^2\;\hat{g}_4}{48\nu(\beta+\hat{g}_2)^2},\ \ \ \Rightarrow\ \ \ \eta\rho+d_t\rho=0\;.
\end{equation}
Finally, we compute the scaling of the terms $\tr\phi^4$. Here we consider any contribution of the forms $\eta \hat{g}_{4}^2$, $d_t\omega \hat{g}_{4}^2$, etc. as negligible, because they would yield higher orders in $\hat{g}_4$ upon substitution. Then, following (\ref{equ:beta-litimR}), we compute the trace for $\hat{g}_4$ and extract the zeroth order in $\hat{g}_2$
\begin{equation}\label{equ:g-zeta}
\mathcal{Z}^{-2}d_t(\mathcal{Z}^2\hat{g}_4)=2\hat{g}_{4}\eta+d_t\hat{g}_4\approx \frac{(\hat{g}_4)^2}{96\pi^2}+O(\hat{g_2},\hat{g}_4^3)\;,
\end{equation}
and similarly the one for the mass, with the quadratic divergence (which depends of the scheme parameter $\beta$)
\begin{equation}\label{equ:m-zeta}
\hat{g}_{2}\eta+d_t\hat{g}_2= -2\hat{g}_2-\frac{\beta\hat{g}_4}{192\pi^2}+\frac{\hat{g}_4\;\hat{g}_2}{96\pi^2}+O(\hat{g_2}^2,\hat{g}_4^2)\;.
\end{equation}
We are then left with four differential equations (\ref{equ:omega-zeta}, \ref{equ:rho-zeta}, \ref{equ:g-zeta}, \ref{equ:m-zeta}) in five variables. This indeterminacy reflects the presence of a redundant coupling, i.e. we can fix the running of any of the couplings by a field redefinition. In this way, we can recover the perturbation theory results, considering only the logarithmic divergences and to the first non-vanishing term in the squared mass $\hat{g}_2$
\begin{align}
&d_t\hat{g}_4=O(\hat{g}_4^3,\frac{1}{\theta^2k^4}),\ \ \ \ &d_t\hat{g}_2=-2\hat{g}_2+\frac{\hat{g}_4\;\hat{g}_2}{192\pi^2}+O(\hat{g}_4^2,\frac{1}{\theta^2k^4}),\\
&\eta=\frac{\hat{g}_4}{192\pi^2}+O(\hat{g}_4^2,\frac{1}{\theta^2k^4}), \ \ \ \ \ &d_t\omega=-\frac{\hat{g}_4\ \omega}{192\pi^2}+O(\hat{g}_4^2,\frac{1}{\theta^2k^4}),
\end{align}
in agreement with \cite{grossewulkenhaar:2004} at the first order in the couplings\footnote{To see this, besides accounting for the aforementioned different definition of the scale derivative and of $\eta$, one must expand $\mathcal{N}\frac{d}{d\mathcal{N}}\Omega=\beta_\Omega$ to the first order in $\Omega=\sqrt{1+\omega}$.}. In particular, we recover the important result that \emph{the self dual model at $\omega=0$ is at one loop an attractive fixed point for the RG flow}. One can improve this approximation by including in the computation the $\beta$-functions coming from $\dot{R}$, which give corrections at higher orders in $\hat{g}_4$.

This choice of the regulator predicts the next to leading order terms in $\omega$ to appear only at higher orders in $\hat{g}_4$, in contrast with the perturbation theory result, which is what we would get using (\ref{equ:regulator-sharp}), and taking the limit $\alpha\to0$. One should not think that (\ref{equ:regulator-covariant}) approximates the perturbative results, but rather that it is a genuinely different regularization scheme. Indeed there is reason to believe that the \lq\lq optimized\rq\rq\ regulator converges faster, when enlarging the truncation, because it effectively accounts for higher loop effects \cite{litim:2001}. As remarked, the particular choice of the regulator is irrelevant when considering the full theory space, where the flow equation is exact.

\section{Conclusions}

We have studied the functional renormalization of the Grosse-Wulkenhaar model, a non-commutative scalar quantum field theory, in its matrix formulation. Using a technical assumption we have shown how the symmetry of the self-dual GW model allows to study the RG flow in the whole theory space, constructed as the smallest functional space containing the bare action which is a priori stable under the renormalization flow. We retrieved the perturbation theory result that the model is asymptotically safe, in the sense that there is a line of fixed point (a one-dimensional UV attractor) in the vicinity of the Gaussian fixed point labeled by a nonvanishing $\phi^4$-coupling.

Our approach relies on a linearization of the flow in the vicinity of the Gaussian fixed-point, and thus cannot be straightforwardly applied to the case of a generic non Gaussian fixed point for large values of the couplings, as it seems to be the case in Gravity. Let us stress two points that may apply to that more complicated scenario. First, it was crucial for us to carefully construct the theory space. This may be the case also in Gravity, where indeed it has been shown how some choices of theory space may spoil the asymptotic safety scenario \cite{machado:2007}. Second, we needed the fact that the mass dimensions of the irrelevant couplings grow at the GFP, and prevent them from giving a collective relevant contribution. A similar feature is desirable also at the NGFP of Gravity, and if present would indicate that high powers of the curvature tensors should receive small quantum corrections to their anomalous dimension. A qualitative argument in this sense was provided by Weinberg in the seminal paper \cite{weinberg:1979}, and the scenario is confirmed for the first few powers of the Ricci scalar \cite{codellopercacci:2007}.

Finally, let us remark that extending the FRGE to the general duality-covariant formalism in a truncation was straightforward, and we could easily retrieve the one loop results of \cite{grossewulkenhaar:2004}. This indicates that the FRGE may be a powerful tool in the investigation of matrix models, and, where a dimensional reasoning similar to Section \ref{sec:beta-function} can be established, perhaps also tensor models such as group field theories.

\subsection*{Acknowledgments}

We would like to give special tanks to Razvan Gurau for sharing many insights on the GW model, for carefully reading the manuscript and being available for countless discussions. We also thank Roberto Percacci for discussions on asymptotic safety. Research at Perimeter Institute is supported by the Government of Canada through Industry Canada and by the Province of Ontario through the Ministry of Research and Innovation. AS gratefully acknowledges financial support by the Istituto Nazionale di Fisica Nucleare, sezione di Padova, and the Scuola Galileiana di Studi Superiori in Padova.

\appendix
\section{Appendix}

\subsection{Identities for Traces}\label{app:traces}

Let us now consider the full theory space (\ref{equ:full-theory-space}), and list the couplings that can be generated from the trace in the r.h.s. of (\ref{equ:vertex-expansion}) which are proportional to $\tr\phi^4$, when we use the \lq\lq optimized\rq\rq\ regulator (\ref{equ:regulator-litim}). Since we must end up with four powers of $\phi$, we can expect dependence on the planar terms $g_{2,\{n\}}$, $g_{4,\{n\}}$ and $g_{6,\{n\}}$ and on the non-planar ones $g_{1,n|1,m}$, $g_{1,n|3,\{m\}}$, $g_{2,\{n\}|2,\{m\}}$, $g_{1,n|5,\{m\}}$, $g_{2,\{n\}|4,\{m\}}$. The contributions arising from contractions of non-planar terms, however, will be suppressed in the UV by powers of $k^{-4}\theta^{-2}$.

Notice that due to the presence of the terms proportional to $g_{1|1}$, $P_{abcd}$ in (\ref{equ:vertex-expansion}) is no longer diagonal, as it includes terms of the form ${\hat\theta}^{-2}\delta_{ab}\delta_{cd}$, that is the projector on the trace part. The only case where we cannot use the cyclic property of the trace to sort all the $P$s to one side is the quadratic term $\dot{R}.P.P.F.P.F$; the inverse $P$ can be expanded in a \lq\lq planar\rq\rq\ part involving only the identity $P_{pl}$ and in a non-planar one $P_{np}$, proportional to the projector ${\hat \theta}^{-2} \delta_{ab}\delta_{cd}$. However, one has that any contraction with the non-planar part of the propagator of the form $F.P_{np}.F.P_{np}$ gives a non-planar term, so we can as well write, in the planar sector,
\begin{equation}\label{equ:planar-projection}
\left.\dot{R}.P.P.F.P.F\right|_{pl}=\left.\dot{R}.P.P.F.P_{pl}.F\right|_{pl}.
\end{equation}
 Since the planar part $P_{pl}$ is diagonal and commutes with any other matrix up to the dependence on a function of the indices in $P$, we can think that the non trivial matrix structure arises just from the product $F.F$. Furthermore, any number of factors $\delta_{ab}\delta_{cd}$ acting on $F.F$ will just act as projectors on the trace, even if the presence of such terms. As for the index dependence in the propagator appearing in $F.P.F$, it can be straightforwardly addressed as in the Section \ref{sec:oneloop-dual}, cfr. (\ref{equ:trace-w-P}); indeed for any planar contribution
\begin{equation}
F.P.F|_{pl}=F_{abcd}\;P(c,d)\delta^{cf}\delta^{de}\;F_{efgh}\delta^{da}\delta^{gh}|_{pl}\approx\delta_{aa}(F^2)_{bbcc} P(a,c),
\end{equation}
so that it is enough to restrict the summation to extract the factor of $P(a,c)$ and evaluate it at $c=0$.

Now let us consider the insertion of operators $K^n$. For instance, the term $\tr [\phi^6K^n]=\tr [\phi_{ab}...\phi_{fa}(\frac{a+1}{\theta})^n]$ will generate a term of the form $(\frac{r+1}{\theta})^n\delta_{qr}(\phi^4)_{sp}$ when the functional derivative $\delta^2/\delta\phi_{pq}\delta\phi_{rs}$ acts exactly on $\phi_{ab}$ and  $\phi_{fa}$. Then, after tracing, one is left with a term proportional to $\tr\phi^4$. Observe that, due to the restriction on where the derivative must act, the eligible terms in the Hessian of $g_{i,n}\tr[\phi^iK^n]$ will be $i$ times fewer than the ones from $g_{i}\tr[\phi^i]$. This reasoning no longer holds when considering, e.g. $\tr [\phi^5K^n\phi K^m]$, because its Hessian can yield only $(\frac{r+1}{\theta})^n\delta_{qr}(\phi^4)_{sp}(p+1)^m$ or $(\frac{s+1}{\theta})^m\delta_{sp}(\phi^4)_{qr}(q+1)^n$, whose traces are proportional to $\tr[\phi^4 K^m]$ and $\tr[\phi^4 K^n]$ respectively; so we can restrict the dependence of the $\beta$-function to $g_{2,n}$, $g_{4,n}$ and $g_{6,n}$.

The same reasoning allows to restrict to non-planar terms of the form $g_{1,n|1}$, $g_{1|3}$, $g_{2|2}$, $g_{1|5}$, $g_{2,n|4}$. For instance, let us consider a term of the form
\begin{equation}
g_{2,k|2}\tr[K^k\phi^2]\tr[\phi^2]=g_{2,k|2}(\frac{a+1}{\theta})^k\phi_{ab}\phi_{ba}\,\phi_{cd}\phi_{dc}.
\end{equation}
To obtain a contribution proportional to $\tr\phi^4$ we must take part of the Hessian where the two derivatives pick a $\phi$ from each trace. This yields $(\frac{q+1}{\theta})^k\phi_{qp}\phi_{sr}$, which can be contracted with a planar contribution:
\begin{equation}
(\frac{q+1}{\theta})^k\phi_{qp}\phi_{sr} \delta^{sa}\delta^{rb}(\phi^2)_{da}\delta_{bc}\;\delta^{pd}\delta^{qc}=\tr[K^k\phi^4],
\end{equation}
so that we must require $k=0$.

Furthermore, terms proportional to $g_{1,n|1}$ can only be contracted with non-planar terms. In fact, in this case one gets
\begin{equation}
(a+1)^k\delta_{ba}\delta_{dc}\delta^{cn}\delta^{dm}\delta_{nm}\phi^4_{qp}\delta^{aq}\delta^{bp},
\end{equation}
which yields a contribution proportional to $\tr\phi^4$, even if suppressed in $\theta^{-1} k^{2}$; on the other hand the replacement with a planar term $\delta_{nm}\phi^4_{qp}\to\delta_{qm}\phi^4_{np}$ would lead to $\tr[K^k\phi^4]$.

The same reasoning can be applied to find the terms appearing in $\eta$. The main difference there is that we can allow for the presence of one \lq\lq spare\rq\rq\ operator $K$, which will end up contributing to $\tr[K\phi^2]$; for a detailed presentation of this point, see Appendix \ref{app:explicit-vertex}.

We conclude this section with a bound that will be useful when computing the traces:
\begin{equation}\label{equ:bound-integral}
\sum_{a_1,a_2}(a+1)^p\,H[C-(a+1)]\leq\int_0^C\!\!\! da_1\int_0^{C-a_1}\!\!\!\!\!\!\!\!\!\!\!\!d a_2\;(a_1+a_2+1)^p\leq  C^{p+2}.
\end{equation}

\subsection{Bounds for the $\beta$-function: an example}\label{app:example}

We want to provide here the full details of one of the calculations occurring in the bounds discussed in Section \ref{sec:full-vertex-exp}. Let us consider the contributions from the r.h.s. of (\ref{equ:vertex-expansion}) proportional to $\hat{g}_{6,n}\tr\phi^4$, which we denote as $\Xi_6$. We introduce the short-hand $\hat{\theta}:=k^2\theta/4$ and, for simplicity, take $\beta=1$ in the \lq\lq optimized\rq\rq\ cutoff function. We also write everything in terms of the dimensionless quantities, denoted by a hat.
\begin{equation}
\Xi_6=-\tilde{C}(1+\eta/2)\sum_{a_1,a_2}^{\hat{\theta}} \frac{\sum_{p=0}^\infty \left(\frac{a+1}{\hat{\theta}}\right)^p\ C_p\hat{g}_{6,p}}{\left(1+\hat{g}_2+\sum_{q=2}^\infty \left(\frac{a+1}{\hat{\theta}}\right)^q\;\hat{g}_{2,q}+\hat{g}_{1|1}\right)^2}.
\end{equation}
\par
{\bf Step 1.} The combinatorial coefficient coming from the Hessian for the terms $g_{6}$ is larger (by a factor of six) than the one for $g_{6,n}$ with $n>0$, because there the derivative has to act on the $\phi$s adjacent to $K^n$, i.e. $C_0/C_n=6$. We overestimate $|\Xi_6|$ setting $C_n=C_0$. For convenience, let us define $C:=\tilde{C}C_0(1+\eta/2)$. By expanding the denominator in an asymptotic series and regrouping powers of $a+1$, we get
\begin{align}\nonumber
|\Xi_6|&\leq C\!\!\sum_{a_1,a_2}^{\hat{\theta}} \left|\sum_{p=0}^{\infty}\! {\scriptstyle \left(\frac{a+1}{\hat{\theta}}\right)^p} \sum_{s=0}^p\!\hat{g}_{6,p-s}\!\!\sum_{k=0}^{s}\frac{k(-1)^k}{\scriptstyle (1+\hat{g}_2+\hat{g}_{1|1})^{k+2}} \prod_{n=0}^k\sum_{i_n=2}^k\!\!\!\hat{g}_{2,i_n}\! \delta\!\!\left(\sum_{n=1}^k i_n -s\right)\right|\\
\nonumber
&\leq\!C\!\!\sum_{a_1,a_2}^{\hat{\theta}} \sum_{p=0}^{\infty}\! {\scriptstyle \left(\frac{a+1}{\hat{\theta}}\right)^p} \sum_{s=0}^p\!|\hat{g}_{6,p-s}|\!\!\sum_{k=0}^{s}\frac{k}{\scriptstyle |1+\hat{g}_2+\hat{g}_{1|1}|^{k+2}} \prod_{n=0}^k\sum_{i_n=2}^k\!|\hat{g}_{2,i_n}| \delta\!\!\left(\sum_{n=1}^k i_n -s\right)\\ \nonumber
&=\!C\!\sum_{p=0}^{\infty}\!\sum_{a_1,a_2}^{\hat{\theta}} \!\! {\scriptstyle \left(\frac{a+1}{\hat{\theta}}\right)^p} \sum_{s=0}^p\!|\hat{g}_{6,p-s}|\!\!\sum_{k=0}^{s}\frac{k}{\scriptstyle |1+\hat{g}_2+\hat{g}_{1|1}|^{k+2}} \prod_{n=0}^k\sum_{i_n=2}^k\!|\hat{g}_{2,i_n}| \delta\!\!\left(\sum_{n=1}^k i_n -s\right)\\
&\leq\!C\!\sum_{p=0}^{\infty}\!M_0^p \sum_{s=0}^p\!|\hat{g}_{6,p-s}|\!\!\sum_{k=0}^{s}\frac{k}{\scriptstyle |1+\hat{g}_2+\hat{g}_{1|1}|^{k+2}} \prod_{n=0}^k\sum_{i_n=2}^k\!|\hat{g}_{2,i_n}| \delta\!\!\left(\sum_{n=1}^k i_n -s\right),
\end{align}
where in the last two passages we changed the order of summation and performed the trace over $a$.

\par
{\bf Step 2.} We straightforwardly have, up to allowing for a possibly larger $C$
\begin{align}
|\Xi_6|\leq\!C\!\sum_{p=0}^{\infty}\!M_2^p \sum_{s=0}^p\!|\hat{g}_{6,p-s}|\prod_{n=0}^k\sum_{i_n=2}^k\!|\hat{g}_{2,i_n}| \delta\!\!\left(\sum_{n=1}^k i_n -s\right).
\end{align}
In particular, observe that we accounted for the factor of $k$ in $M_2^p$, given that $k\leq p$.

\par
{\bf Step 3.} We now proceed to reorder the sum according to the mass dimension of the couplings involved in it. Observe that the zeroth order contribution $g_{6}$ has dimension $-d_o=-2$. Furthermore, when the combined dimension of the coupling is $-2-2p$ they can at most give a factor of $(a+1)^{2p}$, and this happens when $k=s=p$. It follows that it is enough to take $M_3\approx M_2^2$
\begin{align}
\left|\Xi_6\right|&\leq C\!\sum_{q=0}^{\infty} M_3^p\sum_{s=0}^{q}\sum_{k=0}^{s} \prod_{n=0}^k\left[\sum_{i_n=2}^k|\hat{g}_{2,i_n}|\right]|\hat{g}_{6,q-s+k}| \delta\left(\sum_{n=1}^k i_n -s\right).
\end{align}

\par
{\bf Step 4.}  We can now use (\ref{equ:bound-irrelevant}) to find
\begin{equation}
\left|\Xi_6\right|\leq C_2 \sum_{q=0}^{\infty} e^{-\tau(2q-\varepsilon q+2-\varepsilon)+\mu_3 q} \sum_{s=0}^{q}\sum_{k=0}^{s} \prod_{n=0}^k\left[\sum_{i_n=2}^k\right]\;\delta\left(\sum_{n=1}^k i_n -s\right),
\end{equation}
with $\mu_3=\log M_2$. The sums over $i_n$, constrained by the $\delta$-function, count the number of ways to put $s$ objects in $k$ boxes, putting at least two in each box. This is smaller than the number of ways to do so without the last condition, that is $N_{s,k}:=\left(\begin{array}{c}k+s-1\\ s\end{array}\right)$ in term of the binomial coefficient. Then, extracting the scaling due to $d_o$,
\begin{align}\nonumber
\left|\Xi_6\right|&\leq Ce^{(-2+\varepsilon)\tau} \sum_{q=0}^{\infty} e^{-\tau(2q-\varepsilon q)+\mu_3 q} \sum_{s=0}^{p}\sum_{k=0}^{s} N_{s,k}\leq Ce^{(-2+\varepsilon)\tau}\sum_{q=0}^{\infty} e^{-\tau(2q-\varepsilon q)+\mu q}\\
\left|\Xi_6\right|&\lesssim  N\; e^{(-2+\varepsilon)\tau}=\tilde{N}(1+\eta/2)\; e^{(-2+\varepsilon)\tau},
\end{align}
when $\tau$ is large enough, i.e. the neighborhood in the coupling space is small enough. Let us remark that the above estimate is very generous, and in practice one could take $M_2\approx 1$ and end up with $\mu$ of the order of some unity.

\subsection{Vanishing of planar tadpoles with \lq\lq jumps\rq\rq}\label{app:non-planar}

We want to prove that at one loop the flow equation (\ref{equ:wetterich-equation}) cannot generate any term of the form $\tr[\phi\mathcal{H}\phi]$, where $\mathcal{H}$ is the non-diagonal part of the kinetic term, see (\ref{equ:definition-H}). Such a term may only come from $\tr[F\dot{R}]$, more specifically from $\tr[F\mathcal{H}]$. Observe that $\mathcal{H}$ is the sum of four terms; let us set, for brevity
\begin{equation}
\sqrt{mn}\delta_{m-1,l}\delta_{n-1,k}:=\sqrt{m_1n_1}\delta_{m_1-1,l_1}\delta_{n_1-1,k_1}\delta_{m_2,l_2}\delta_{n_2,k_2}+\{1\}\leftrightarrow\{2\},
\end{equation}
and similarly for $\sqrt{(m+1)(n+1)}\delta_{m+1,l}\delta_{n+1,k}$. It follows, up to numerical coefficients
\begin{align}\nonumber
\tr\left[F\mathcal{H}\right]\approx&\left(\phi_{da}\phi_{bc}+\phi^2_{da}\delta_{bc}+\phi^2_{bc}\delta_{da}\right)\;\delta^{cn}\delta^{md}\;\delta^{al}\delta^{bk}\\
&\left(\sqrt{mn}\delta_{m-1,l}\delta_{n-1,k}+\sqrt{(m+1)(n+1)}\delta_{m+1,l}\delta_{n+1,k}\right).
\end{align}
For instance, the contraction of the $\phi\phi$ term gives 
\begin{equation}
\tr\left[\sqrt{m}\phi_{m,m-1}\right]\tr\left[\sqrt{n}\phi_{n-1,n}\right]=\sum_{m,n}\sqrt{m_1n_1}\phi^{m_2,m_2}_{m_1-1,m_1-1}\phi^{n_2,n_2}_{n_1-1,n_1}+\{1\}\leftrightarrow\{2\},
\end{equation}
that is a non-planar term. On the other hand, contracting any of the $\phi^2\delta$ terms yields a contribution proportional to $\delta_{n\pm1,n}$, which therefore vanishes.

\begin{adjustwidth}{-1cm}{-1cm}

\section{Explicit Vertex Expansion}\label{app:explicit-vertex}

This appendix provides the explicit form of the terms needed in the vertex expansion of the self-dual model:
\begin{equation}\label{equ:vertex-exp}
  \Gamma[\phi]=\sum_{i=1}^{\infty}\Gamma_{2i}[\phi]=\sum_{i=1}^\infty G_{2i}^{m_1n_1...m_{2i}n_{2i}}\phi_{m_1n_1}...\phi_{m_{2i}n_{2i}},
\end{equation}
which implies the Hessian
\begin{equation}
  \Gamma^{(2)abcd}[\phi]=\sum_{i=1}^\infty F_{2i}^{abcdm_3n_3...m_{2i}n_{2i}}\phi_{m_3n_3}...\phi_{m_{2i}n_{2i}},
\end{equation}
where
\begin{equation}
  \begin{array}{rcl}
    F_{2i}^{abcdm_3n_3...m_{2i}n_{2i}}&=&G_{2i}^{abcd...m_{2i}n_{2i}}+...+G_{2i}^{ab...cd...m_{2i}n_{2i}}+...+G_{2i}^{abm_3n_3...cd}\\
      &+&...\\
      &+&G_{2i}^{cd...ab...m_{2i}n_{2i}}+...+G_{2i}^{m_3n_3...ab...cd}\\
      &+&...\\
      &+&G_{2i}^{cd...ab}+...+G_{2i}^{m_3n_3...cd...ab}+...+G_{2i}^{m_3n_3...dcab}.
  \end{array}
\end{equation}
We will now investigate which field monomials can contribute to the running of the couplings $Z,m^2,\lambda$ in the self-dual GW-action form the first terms in the vertex expansion, where we denote the inverse of the field independent part of $\Gamma^{(2)}+R$ by $P$:
\begin{equation}\label{equ:explicit-vertex-expansion}
  \begin{array}{rcl}
    \partial_t \Gamma_k[\phi]&=& \frac 1 2 \dot R_k^{abcd}\left(P_{abcd}\right.\\
                             &&-P_{abr_1s_1}F^{r_1s_1r_2s_2m_1n_1m_2n_2}P_{r_2s_2cd}\phi_{m_1n_1}\phi_{m_2n_2}\\
                             &&+\left(P_{abr_1s_1}F^{r_1s_1r_2s_2m_1n_1m_2n_2}P_{abr_1s_1}F^{r_1s_1r_2s_2m_1n_1m_2n_2}P_{r_2s_2cd}\right.\\
                             &&\left.\left.-P_{abr_1s_1}F^{r_1s_1r_2s_2m_1n_1m_2n_2m_3n_3m_4n_24}P_{r_2s_2cd}\right)\phi_{m_1n_1}...\phi_{m_4n_4}\right)\\
                             &&+\mathcal O(\phi^8).
  \end{array}
\end{equation}
Here and throughout we will discard the first term as a pure vacuum term.

{\it First,} only $\Gamma_{2}[\phi],\Gamma_4[\phi]$ and $\Gamma_6[\phi]$ contribute to the running of the couplings in the GW-action, since the Hessian of all other terms contains at least six fields. In this Appendix we will use a slightly different, self explanatory notation for the couplings, which is more compact when it comes to write down the vertex expansion. It is convenient to expand these in terms of their mass dimension:\begin{equation}
  \Gamma_2[\phi]= g_{2,(n_1,n_2)}[\phi K^{n_1} \phi K^{n_2}]+ g_{2,(n),(m)}[\phi K^n][\phi K^m],
\end{equation}
\begin{equation}
  \begin{array}{rcl}
    \Gamma_4[\phi]&=&  g_{4,(n_1,...,n_4}[\phi K^{n_1}\phi K^{n_2}\phi K^{n_3}\phi K^{n_4}]\\
                  &+&  g_{4,(n_1,n_2,n_3),(m)}[\phi K^{n_1}\phi K^{n_2}\phi K^{n_3}][\phi K^m]\\
                  &+&  g_{4,(n_1,n_2),(m_1,m_2)}[\phi K^{n_1}\phi K^{n_2}][\phi K^{m_1}\phi K^{m_2}]\\
                  &+&  g_{4,(n_1,n_2),(m),(o)}[\phi K^{n_1}\phi K^{n_2}][\phi K^m][\phi K^o]]\\
                  &+&  g_{4,(n),(m),(o),(p)}[\phi K^{n}][\phi K^{m}][\phi K^{o}][\phi K^{p}],
  \end{array}
\end{equation}
\begin{equation}
  \begin{array}{rcl}
    \Gamma_6[\phi]&=& g_{6,(n_1,n_2,n_3,n_4,n_5,n_6)}[\phi K^{n_1}\phi K^{n_2}\phi K^{n_3}\phi K^{n_4}\phi K^{n_5}\phi K^{n_6}]\\
                  &+& g_{6,(n_1,n_2,n_3,n_4,n_5),(m)}[\phi K^{n_1}\phi K^{n_2}\phi K^{n_3}\phi K^{n_4}\phi K^{n_5}][\phi K^{m}]\\
                  &+& g_{6,(n_1,n_2,n_3,n_4),(m_1,m_2)}[\phi K^{n_1}\phi K^{n_2}\phi K^{n_3}\phi K^{n_4}][\phi K^{m_1}\phi K^{m_2}]\\
                  &+& g_{6,(n_1,n_2,n_3),(m_1,m_2,m_3)}[\phi K^{n_1}\phi K^{n_2}\phi K^{n_3}][\phi K^{m_1}\phi K^{m_2}\phi K^{m_3}]\\
                  &+& g_{6,(n_1,n_2,n_3,n_4),(m),(o)}[\phi K^{n_1}\phi K^{n_2}\phi K^{n_3}\phi K^{n_4}][\phi K^{m}][\phi K^{o}]\\
                  &+& g_{6,(n_1,n_2,n_3),(m_1,m_2),(o)}[\phi K^{n_1}\phi K^{n_2}\phi K^{n_3}][\phi K^{m_1}\phi K^{m_2}][\phi K^{o}]\\
                  &+& g_{6,(n_1,n_2),(m_1,m_2),(o_1,o_2)}[\phi K^{n_1}\phi K^{n_2}][\phi K^{m_1}\phi K^{m_2}][\phi K^{o_1}\phi K^{o_2}]\\
                  &+& g_{6,(n_1,n_2,n_3),(m),(o),(p)}[\phi K^{n_1}\phi K^{n_2}\phi K^{n_3}][\phi K^{m}][\phi K^{o}][\phi K^{p}]\\
                  &+& g_{6,(n_1,n_2),(m_1,m_2),(o),(p)}[\phi K^{n_1}\phi K^{n_2}][\phi K^{m_1}\phi K^{m_2}][\phi K^{o}][\phi K^{p}]\\
                  &+& g_{6,(n_1,n_2),(m),(o),(p),(q)}[\phi K^{n_1}\phi K^{n_2}][\phi K^{m}][\phi K^{o}][\phi K^{p}][\phi K^{q}]\\
                  &+& g_{6,(n),(m),(o),(p),(q),(r)}[\phi K^{n}][\phi K^{m}][\phi K^{o}][\phi K^{p}][\phi K^{q}][\phi K^{r}],
  \end{array}
\end{equation}
where square brackets on the r.h.s. denote traces and the appropriate powers of $\theta$ (see section \ref{sec:theory-space}). Repeated indices are summed over and where we caution that not all coupling constants are independent due to cyclicity of the trace and commutativity of the product of traces. These redundancies can be resolved by restricting the sums to obey $i_1\ge i_j$ in each round bracket and $i_1\ge j_1$ if $i_k$ appear in a round bracket of same length left of the round bracket containing $j_l$ and selecting one representative in the special cases of monomials with symmetries.

{\it Second,} only terms whose Hessian contains at most one trace over fields can contribute to the running of couplings in the GW-action, since the monomials in the GW-action contain only one trace, hence we have
\begin{equation}
  \begin{array}{rcl}
    \Gamma_4[\phi]-R_4[\phi]&=& g_{4,(n_1,...,n_4}[\phi K^{n_1}\phi K^{n_2}\phi K^{n_3}\phi K^{n_4}]\\
                  &+& g_{4,(n_1,n_2,n_3),(m)}[\phi K^{n_1}\phi K^{n_2}\phi K^{n_3}][\phi K^m]\\
                  &+& g_{4,(n_1,n_2),(m_1,m_2)}[\phi K^{n_1}\phi K^{n_2}][\phi K^{m_1}\phi K^{m_2}]\\
                  &+& g_{4,(n_1,n_2),(m),(o)}[\phi K^{n_1}\phi K^{n_2}][\phi K^m][\phi K^o]],
  \end{array}
\end{equation}
\begin{equation}
  \begin{array}{rcl}
    \Gamma_6[\phi]-R_6[\phi]&=&g_{6,(n_1,n_2,n_3,n_4,n_5,n_6)}[\phi K^{n_1}\phi K^{n_2}\phi K^{n_3}\phi K^{n_4}\phi K^{n_5}\phi K^{n_6}]\\
                  &+& g_{6,(n_1,n_2,n_3,n_4,n_5),(m)}[\phi K^{n_1}\phi K^{n_2}\phi K^{n_3}\phi K^{n_4}\phi K^{n_5}][\phi K^{m}]\\
                  &+& g_{6,(n_1,n_2,n_3,n_4),(m_1,m_2)}[\phi K^{n_1}\phi K^{n_2}\phi K^{n_3}\phi K^{n_4}][\phi K^{m_1}\phi K^{m_2}]\\
                  &+& g_{6,(n_1,n_2,n_3,n_4),(m),(o)}[\phi K^{n_1}\phi K^{n_2}\phi K^{n_3}\phi K^{n_4}][\phi K^{m}][\phi K^{o}],
  \end{array}
\end{equation}
where $R_4[\phi]$ and$R_6[\phi]$ can be neglected. Let us denote these summands by $\Gamma_4[\phi]-R_4[\phi]=\sum_{i=1}^4\Gamma_{4,i}[\phi]$ and $\Gamma_6[\phi]-R_6[\phi]=\sum_{i=1}^4 \Gamma_{6,i}[\phi]$ and investigate their Hessians:
\begin{equation}\label{equ:gamma-2-2}
  \begin{array}{rcl}
    \Gamma_2[\phi]^{rstu}&=& g_{2,(n_1,n_2)}\left(K^{n_1}_{s,t}K^{n_2}_{u,r}+K^{n_2}_{s,t}K^{n_2}_{u,r}\right)\\
                         &+& g_{2,(n),(m)}\left(K^n_{r,s}K^m_{t,u}+K^m_{r,s}K^n_{t,u}\right).
  \end{array}
\end{equation}
\begin{equation}
  \begin{array}{rcl}
    \Gamma_{41}^{rstu}&=& g_{4,(n_1,n_2,n_3,n_4)}\phi_{a_1b_1}\phi_{a_2b_2}\left(
    K_{u,a_1}^{n_2} K_{b_1,a_2}^{n_3} K_{b_2,r}^{n_4} K_{s,t}^{n_1}
    +K_{u,a_2}^{n_3} K_{b_1,r}^{n_1} K_{b_2,a_1}^{n_4} K_{s,t}^{n_2}\right.\\
    &+&K_{u,a_1}^{n_4} K_{b_1,a_2}^{n_1}K_{b_2,r}^{n_2} K_{s,t}^{n_3}
   +K_{u,a_1}^{n_1} K_{b_1,a_2}^{n_2} K_{b_2,r}^{n_3} K_{s,t}^{n_4}+K_{s,a_1}^{n_4} K_{u,r}^{n_3} K_{b_1,a_2}^{n_1}
   K_{b_2,t}^{n_2}\\
   &+&K_{s,a_1}^{n_4} K_{u,a_2}^{n_2} K_{b_1,t}^{n_1} K_{b_2,r}^{n_3}+K_{s,a_2}^{n_2} K_{u,a_1}^{n_4} K_{b_1,r}^{n_1} K_{b_2,t}^{n_3}+K_{s,a_1}^{n_1}
   K_{u,r}^{n_4} K_{b_1,a_2}^{n_2} K_{b_2,t}^{n_3}\\
   &+&K_{s,a_1}^{n_1} K_{u,a_2}^{n_3} K_{b_1,t}^{n_2} K_{b_2,r}^{n_4}+K_{s,a_2}^{n_3} K_{u,a_1}^{n_1} K_{b_1,r}^{n_2}
   K_{b_2,t}^{n_4}+K_{s,a_1}^{n_2} K_{u,r}^{n_1} K_{b_1,a_2}^{n_3} K_{b_2,t}^{n_4}\\
   &+&\left.K_{s,a_2}^{n_3} K_{u,r}^{n_2} K_{b_1,t}^{n_1} K_{b_2,a_1}^{n_4}\right)
  \end{array}
\end{equation}
\begin{equation}
  \begin{array}{rcl}
    \Gamma_{42}^{rstu}&=&  g_{4,(n_1,n_2,n_3),(m)}\phi_{a_1b_1}\phi_{a_2b_2}\left(K_{u,a_1}^{n_3} K_{b_1,a_2}^{n_1} K_{b_2,t}^{n_2} K_{s,r}^m
    +K_{u,a_1}^{n_1} K_{b_1,a_2}^{n_2} K_{b_2,t}^{n_3} K_{s,r}^m\right.\\
    &+&K_{u,a_2}^{n_2} K_{b_1,t}^{n_1}
   K_{b_2,a_1}^{n_3} K_{s,r}^m+K_{s,t}^{n_2} K_{u,a_1}^{n_3} K_{b_1,r}^{n_1} K_{b_2,a_2}^m
   +K_{s,a_1}^{n_3} K_{u,r}^{n_2} K_{b_1,t}^{n_1} K_{b_2,a_2}^m\\&+&K_{s,t}^{n_3}
   K_{u,a_1}^{n_1} K_{b_1,r}^{n_2} K_{b_2,a_2}^m+K_{s,a_1}^{n_1} K_{u,r}^{n_3} K_{b_1,t}^{n_2} K_{b_2,a_2}^m
   +K_{s,t}^{n_1} K_{u,a_1}^{n_2} K_{b_1,r}^{n_3}
   K_{b_2,a_2}^m\\&+&K_{s,a_1}^{n_2} K_{u,r}^{n_1} K_{b_1,t}^{n_3} K_{b_2,a_2}^m+K_{s,a_1}^{n_3} K_{u,t}^m K_{b_1,a_2}^{n_1} K_{b_2,r}^{n_2}
   +K_{s,a_1}^{n_1} K_{u,t}^m
   K_{b_1,a_2}^{n_2} K_{b_2,r}^{n_3}\\&+&\left.K_{s,a_2}^{n_2} K_{u,t}^m K_{b_1,r}^{n_1} K_{b_2,a_1}^{n_3}\right)
  \end{array}
\end{equation}
\begin{equation}
  \begin{array}{rcl}
    \Gamma_{43}^{rstu}&=& g_{4,(n_1,n_2),(m_1,m_2)}\phi_{a_1b_1}\phi_{a_2b_2}\left(K_{u,r}^{m_2} K_{b_1,a_2}^{n_1} K_{b_2,a_1}^{n_2} K_{s,t}^{m_1}+K_{u,r}^{m_1} K_{b_1,a_2}^{n_1} K_{b_2,a_1}^{n_2} K_{s,t}^{m_2}\right.\\
    &+&K_{u,r}^{n_2} K_{b_1,a_2}^{m_1}
   K_{b_2,a_1}^{m_2} K_{s,t}^{n_1}+K_{u,r}^{n_1} K_{b_1,a_2}^{m_1} K_{b_2,a_1}^{m_2} K_{s,t}^{n_2}+K_{s,a_2}^{m_2} K_{u,a_1}^{n_2} K_{b_1,t}^{n_1}
   K_{b_2,r}^{m_1}\\
   &+&K_{s,a_2}^{m_2} K_{u,a_1}^{n_1} K_{b_1,t}^{n_2} K_{b_2,r}^{m_1}+K_{s,a_1}^{n_2} K_{u,a_2}^{m_2} K_{b_1,r}^{n_1} K_{b_2,t}^{m_1}+K_{s,a_1}^{n_1}
   K_{u,a_2}^{m_2} K_{b_1,r}^{n_2} K_{b_2,t}^{m_1}\\
   &+&K_{s,a_2}^{m_1} K_{u,a_1}^{n_2} K_{b_1,t}^{n_1} K_{b_2,r}^{m_2}+K_{s,a_2}^{m_1} K_{u,a_1}^{n_1} K_{b_1,t}^{n_2}
   K_{b_2,r}^{m_2}+K_{s,a_1}^{n_2} K_{u,a_2}^{m_1} K_{b_1,r}^{n_1} K_{b_2,t}^{m_2}\\
   &+&\left.K_{s,a_1}^{n_1} K_{u,a_2}^{m_1} K_{b_1,r}^{n_2} K_{b_2,t}^{m_2}
    \right)
  \end{array}
\end{equation}
\begin{equation}
  \begin{array}{rcl}
    \Gamma_{44}^{rstu}&=& g_{4,(n_1,n_2),(m),(o)}\phi_{a_1b_1}\phi_{a_2b_2}\left(K_{u,a_1}^{n_2} K_{b_1,t}^{n_1} K_{b_2,a_2}^o K_{s,r}^m+K_{u,a_1}^{n_1} K_{b_1,t}^{n_2} K_{b_2,a_2}^o K_{s,r}^m\right.\\
    &+&K_{u,t}^o K_{b_1,a_2}^{n_1} K_{b_2,a_1}^{n_2} K_{s,r}^m+K_{u,a_1}^{n_2} K_{b_1,t}^{n_1} K_{b_2,a_2}^m K_{s,r}^o+K_{u,a_1}^{n_1} K_{b_1,t}^{n_2} K_{b_2,a_2}^m K_{s,r}^o\\
    &+&K_{u,t}^m K_{b_1,a_2}^{n_1}
   K_{b_2,a_1}^{n_2} K_{s,r}^o+K_{s,a_1}^{n_2} K_{u,t}^o K_{b_1,r}^{n_1} K_{b_2,a_2}^m+K_{s,a_1}^{n_1} K_{u,t}^o K_{b_1,r}^{n_2} K_{b_2,a_2}^m\\
   &+&K_{s,t}^{n_2}
   K_{u,r}^{n_1} K_{b_1,a_1}^m K_{b_2,a_2}^o+K_{s,t}^{n_1} K_{u,r}^{n_2} K_{b_1,a_1}^m K_{b_2,a_2}^o+K_{s,a_1}^{n_2} K_{u,t}^m K_{b_1,r}^{n_1}
   K_{b_2,a_2}^o\\
   &+&\left.K_{s,a_1}^{n_1} K_{u,t}^m K_{b_1,r}^{n_2} K_{b_2,a_2}^o
    \right)
  \end{array}
\end{equation}
\begin{equation}
  \begin{array}{rcl}
    \Gamma_{61}^{rstu}&=& g_{6,(n_1,n_2,n_3,n_4,n_5,n_6)}\phi_{a_1b_1}\phi_{a_2b_2}\phi_{a_3b_3}\phi_{a_4b_4}\left(K_{u,a_4}^{n_2} K_{b_1,a_2}^{n_4} K_{b_2,a_3}^{n_5} K_{b_3,r}^{n_6} K_{b_4,a_1}^{n_3} K_{s,t}^{n_1}\right.\\
    &+&K_{u,a_1}^{n_3} K_{b_1,a_2}^{n_4} K_{b_2,a_3}^{n_5}
   K_{b_3,a_4}^{n_6} K_{b_4,r}^{n_1} K_{s,t}^{n_2}+K_{u,a_2}^{n_4} K_{b_1,r}^{n_2} K_{b_2,a_3}^{n_5} K_{b_3,a_4}^{n_6} K_{b_4,a_1}^{n_1} K_{s,t}^{n_3}\\
   &+&K_{u,a_3}^{n_5}
   K_{b_1,a_2}^{n_2} K_{b_2,r}^{n_3} K_{b_3,a_4}^{n_6} K_{b_4,a_1}^{n_1} K_{s,t}^{n_4}+K_{u,a_4}^{n_6} K_{b_1,a_2}^{n_2} K_{b_2,a_3}^{n_3} K_{b_3,r}^{n_4}
   K_{b_4,a_1}^{n_1} K_{s,t}^{n_5}\\
   &+&K_{u,a_4}^{n_1} K_{b_1,a_2}^{n_3} K_{b_2,a_3}^{n_4} K_{b_3,r}^{n_5} K_{b_4,a_1}^{n_2} K_{s,t}^{n_6}+K_{s,a_1}^{n_2} K_{u,a_4}^{n_6}
   K_{b_1,a_2}^{n_3} K_{b_2,a_3}^{n_4} K_{b_3,t}^{n_5} K_{b_4,r}^{n_1}\\
   &+&K_{s,a_1}^{n_2} K_{u,a_3}^{n_5} K_{b_1,a_2}^{n_3} K_{b_2,t}^{n_4} K_{b_3,a_4}^{n_6}
   K_{b_4,r}^{n_1}
   +K_{s,a_1}^{n_2} K_{u,a_2}^{n_4} K_{b_1,t}^{n_3} K_{b_2,a_3}^{n_5} K_{b_3,a_4}^{n_6} K_{b_4,r}^{n_1}\\
   &+&K_{s,a_4}^{n_6} K_{u,a_1}^{n_2}
   K_{b_1,a_2}^{n_3} K_{b_2,a_3}^{n_4} K_{b_3,r}^{n_5} K_{b_4,t}^{n_1}
   +K_{s,a_3}^{n_5} K_{u,a_1}^{n_2} K_{b_1,a_2}^{n_3} K_{b_2,r}^{n_4} K_{b_3,a_4}^{n_6}
   K_{b_4,t}^{n_1}\\
   &+&K_{s,a_2}^{n_4} K_{u,a_1}^{n_2} K_{b_1,r}^{n_3} K_{b_2,a_3}^{n_5} K_{b_3,a_4}^{n_6} K_{b_4,t}^{n_1}
   +K_{s,a_1}^{n_3} K_{u,r}^{n_2} K_{b_1,a_2}^{n_4}
   K_{b_2,a_3}^{n_5} K_{b_3,a_4}^{n_6} K_{b_4,t}^{n_1}\\
   &+&K_{s,a_4}^{n_6} K_{u,r}^{n_5} K_{b_1,a_2}^{n_2} K_{b_2,a_3}^{n_3} K_{b_3,t}^{n_4}
   K_{b_4,a_1}^{n_1}
   +K_{s,a_4}^{n_6} K_{u,a_3}^{n_4} K_{b_1,a_2}^{n_2} K_{b_2,t}^{n_3} K_{b_3,r}^{n_5} K_{b_4,a_1}^{n_1}\\
   &+&K_{s,a_4}^{n_6} K_{u,a_2}^{n_3}
   K_{b_1,t}^{n_2} K_{b_2,a_3}^{n_4} K_{b_3,r}^{n_5} K_{b_4,a_1}^{n_1}
   +K_{s,a_3}^{n_4} K_{u,a_4}^{n_6} K_{b_1,a_2}^{n_2} K_{b_2,r}^{n_3} K_{b_3,t}^{n_5}
   K_{b_4,a_1}^{n_1}\\
   &+&K_{s,a_2}^{n_3} K_{u,a_4}^{n_6} K_{b_1,r}^{n_2} K_{b_2,a_3}^{n_4} K_{b_3,t}^{n_5} K_{b_4,a_1}^{n_1}+K_{s,a_3}^{n_5} K_{u,r}^{n_4}
   K_{b_1,a_2}^{n_2} K_{b_2,t}^{n_3} K_{b_3,a_4}^{n_6} K_{b_4,a_1}^{n_1}\\
   &+&K_{s,a_3}^{n_5} K_{u,a_2}^{n_3} K_{b_1,t}^{n_2} K_{b_2,r}^{n_4} K_{b_3,a_4}^{n_6}
   K_{b_4,a_1}^{n_1}
   +K_{s,a_2}^{n_3} K_{u,a_3}^{n_5} K_{b_1,r}^{n_2} K_{b_2,t}^{n_4} K_{b_3,a_4}^{n_6} K_{b_4,a_1}^{n_1}\\
   &+&K_{s,a_2}^{n_4} K_{u,r}^{n_3} K_{b_1,t}^{n_2}
   K_{b_2,a_3}^{n_5} K_{b_3,a_4}^{n_6} K_{b_4,a_1}^{n_1}
   +K_{s,a_1}^{n_3} K_{u,a_4}^{n_1} K_{b_1,a_2}^{n_4} K_{b_2,a_3}^{n_5} K_{b_3,t}^{n_6}
   K_{b_4,r}^{n_2}\\
   &+&K_{s,a_4}^{n_1} K_{u,a_1}^{n_3} K_{b_1,a_2}^{n_4} K_{b_2,a_3}^{n_5} K_{b_3,r}^{n_6} K_{b_4,t}^{n_2}
   +K_{s,a_4}^{n_1} K_{u,r}^{n_6} K_{b_1,a_2}^{n_3}
   K_{b_2,a_3}^{n_4} K_{b_3,t}^{n_5} K_{b_4,a_1}^{n_2}\\
   &+&K_{s,a_4}^{n_1} K_{u,a_3}^{n_5} K_{b_1,a_2}^{n_3} K_{b_2,t}^{n_4} K_{b_3,r}^{n_6}
   K_{b_4,a_1}^{n_2}
   +K_{s,a_4}^{n_1} K_{u,a_2}^{n_4} K_{b_1,t}^{n_3} K_{b_2,a_3}^{n_5} K_{b_3,r}^{n_6} K_{b_4,a_1}^{n_2}\\
   &+&K_{s,a_3}^{n_5} K_{u,a_4}^{n_1}
   K_{b_1,a_2}^{n_3} K_{b_2,r}^{n_4} K_{b_3,t}^{n_6} K_{b_4,a_1}^{n_2}
   +K_{s,a_2}^{n_4} K_{u,a_4}^{n_1} K_{b_1,r}^{n_3} K_{b_2,a_3}^{n_5} K_{b_3,t}^{n_6}
   K_{b_4,a_1}^{n_2}\\
   &+&\left.K_{s,a_4}^{n_2} K_{u,r}^{n_1} K_{b_1,a_2}^{n_4} K_{b_2,a_3}^{n_5} K_{b_3,t}^{n_6} K_{b_4,a_1}^{n_3}
    \right)
  \end{array}
\end{equation}
\begin{equation}
  \begin{array}{rcl}
    \Gamma_{62}^{rstu}&=& g_{6,(n_1,n_2,n_3,n_4,n_5),(m)}\phi_{a_1b_1}\phi_{a_2b_2}\phi_{a_3b_3}\phi_{a_4b_4}\left(
    K_{u,a_1}^{n_2} K_{b_1,a_2}^{n_3} K_{b_2,a_3}^{n_4} K_{b_3,a_4}^{n_5} K_{b_4,t}^{n_1} K_{s,r}^m\right.\\
    &+&K_{u,a_4}^{n_5} K_{b_1,a_2}^{n_2} K_{b_2,a_3}^{n_3} K_{b_3,t}^{n_4}
   K_{b_4,a_1}^{n_1} K_{s,r}^m+K_{u,a_3}^{n_4} K_{b_1,a_2}^{n_2} K_{b_2,t}^{n_3} K_{b_3,a_4}^{n_5} K_{b_4,a_1}^{n_1} K_{s,r}^m\\
   &+&K_{u,a_2}^{n_3} K_{b_1,t}^{n_2}
   K_{b_2,a_3}^{n_4} K_{b_3,a_4}^{n_5} K_{b_4,a_1}^{n_1} K_{s,r}^m+K_{u,a_4}^{n_1} K_{b_1,a_2}^{n_3} K_{b_2,a_3}^{n_4} K_{b_3,t}^{n_5} K_{b_4,a_1}^{n_2}
   K_{s,r}^m\\
   &+&K_{s,a_1}^{n_2} K_{u,a_4}^{n_5} K_{b_1,a_2}^{n_3} K_{b_2,t}^{n_4} K_{b_3,a_3}^m K_{b_4,r}^{n_1}+K_{s,a_1}^{n_2} K_{u,a_2}^{n_4} K_{b_1,t}^{n_3}
   K_{b_2,a_4}^{n_5} K_{b_3,a_3}^m K_{b_4,r}^{n_1}\\
   &+&K_{s,t}^{n_2} K_{u,a_1}^{n_3} K_{b_1,a_2}^{n_4} K_{b_2,a_4}^{n_5} K_{b_3,a_3}^m K_{b_4,r}^{n_1}+K_{s,a_1}^{n_2}
   K_{u,t}^m K_{b_1,a_2}^{n_3} K_{b_2,a_3}^{n_4} K_{b_3,a_4}^{n_5} K_{b_4,r}^{n_1}\\
   &+&K_{s,a_4}^{n_5} K_{u,a_1}^{n_2} K_{b_1,a_2}^{n_3} K_{b_2,r}^{n_4} K_{b_3,a_3}^m
   K_{b_4,t}^{n_1}+K_{s,a_2}^{n_4} K_{u,a_1}^{n_2} K_{b_1,r}^{n_3} K_{b_2,a_4}^{n_5} K_{b_3,a_3}^m K_{b_4,t}^{n_1}\\
   &+&K_{s,a_1}^{n_3} K_{u,r}^{n_2} K_{b_1,a_2}^{n_4}
   K_{b_2,a_4}^{n_5} K_{b_3,a_3}^m K_{b_4,t}^{n_1}+K_{s,t}^{n_4} K_{u,a_4}^{n_5} K_{b_1,a_2}^{n_2} K_{b_2,r}^{n_3} K_{b_3,a_3}^m K_{b_4,a_1}^{n_1}\\
   &+&K_{s,a_4}^{n_5}
   K_{u,r}^{n_4} K_{b_1,a_2}^{n_2} K_{b_2,t}^{n_3} K_{b_3,a_3}^m K_{b_4,a_1}^{n_1}+K_{s,a_4}^{n_5} K_{u,a_2}^{n_3} K_{b_1,t}^{n_2} K_{b_2,r}^{n_4} K_{b_3,a_3}^m
   K_{b_4,a_1}^{n_1}\\
   &+&K_{s,a_2}^{n_3} K_{u,a_4}^{n_5} K_{b_1,r}^{n_2} K_{b_2,t}^{n_4} K_{b_3,a_3}^m K_{b_4,a_1}^{n_1}+K_{s,t}^{n_3} K_{u,a_2}^{n_4} K_{b_1,r}^{n_2}
   K_{b_2,a_4}^{n_5} K_{b_3,a_3}^m K_{b_4,a_1}^{n_1}\\
   &+&K_{s,a_2}^{n_4} K_{u,r}^{n_3} K_{b_1,t}^{n_2} K_{b_2,a_4}^{n_5} K_{b_3,a_3}^m K_{b_4,a_1}^{n_1}+K_{s,a_4}^{n_5}
   K_{u,t}^m K_{b_1,a_2}^{n_2} K_{b_2,a_3}^{n_3} K_{b_3,r}^{n_4} K_{b_4,a_1}^{n_1}\\
   &+&K_{s,a_3}^{n_4} K_{u,t}^m K_{b_1,a_2}^{n_2} K_{b_2,r}^{n_3} K_{b_3,a_4}^{n_5}
   K_{b_4,a_1}^{n_1}+K_{s,a_2}^{n_3} K_{u,t}^m K_{b_1,r}^{n_2} K_{b_2,a_3}^{n_4} K_{b_3,a_4}^{n_5} K_{b_4,a_1}^{n_1}\\
   &+&K_{s,a_1}^{n_3} K_{u,a_4}^{n_1} K_{b_1,a_2}^{n_4}
   K_{b_2,t}^{n_5} K_{b_3,a_3}^m K_{b_4,r}^{n_2}+K_{s,a_4}^{n_1} K_{u,a_1}^{n_3} K_{b_1,a_2}^{n_4} K_{b_2,r}^{n_5} K_{b_3,a_3}^m K_{b_4,t}^{n_2}\\
   &+&K_{s,t}^{n_5}
   K_{u,a_4}^{n_1} K_{b_1,a_2}^{n_3} K_{b_2,r}^{n_4} K_{b_3,a_3}^m K_{b_4,a_1}^{n_2}+K_{s,a_4}^{n_1} K_{u,r}^{n_5} K_{b_1,a_2}^{n_3} K_{b_2,t}^{n_4} K_{b_3,a_3}^m
   K_{b_4,a_1}^{n_2}\\
   &+&K_{s,a_4}^{n_1} K_{u,a_2}^{n_4} K_{b_1,t}^{n_3} K_{b_2,r}^{n_5} K_{b_3,a_3}^m K_{b_4,a_1}^{n_2}+K_{s,a_2}^{n_4} K_{u,a_4}^{n_1} K_{b_1,r}^{n_3}
   K_{b_2,t}^{n_5} K_{b_3,a_3}^m K_{b_4,a_1}^{n_2}\\
   &+&K_{s,a_4}^{n_1} K_{u,t}^m K_{b_1,a_2}^{n_3} K_{b_2,a_3}^{n_4} K_{b_3,r}^{n_5} K_{b_4,a_1}^{n_2}+K_{s,t}^{n_1}
   K_{u,a_4}^{n_2} K_{b_1,a_2}^{n_4} K_{b_2,r}^{n_5} K_{b_3,a_3}^m K_{b_4,a_1}^{n_3}\\
   &+&\left.K_{s,a_4}^{n_2} K_{u,r}^{n_1} K_{b_1,a_2}^{n_4} K_{b_2,t}^{n_5} K_{b_3,a_3}^m
   K_{b_4,a_1}^{n_3}\right)
  \end{array}
\end{equation}
\begin{equation}
  \begin{array}{rcl}
    \Gamma_{63}^{rstu}&=& g_{6,(n_1,n_2,n_3,n_4),(m_1,m_2)}\phi_{a_1b_1}\phi_{a_2b_2}\phi_{a_3b_3}\phi_{a_4b_4}\left(
    K_{u,r}^{m_2} K_{b_1,a_2}^{n_2} K_{b_2,a_3}^{n_3} K_{b_3,a_4}^{n_4} K_{b_4,a_1}^{n_1} K_{s,t}^{m_1}\right.\\
    &+&K_{u,r}^{m_1} K_{b_1,a_2}^{n_2} K_{b_2,a_3}^{n_3}
   K_{b_3,a_4}^{n_4} K_{b_4,a_1}^{n_1} K_{s,t}^{m_2}+K_{u,a_4}^{n_2} K_{b_1,r}^{n_4} K_{b_2,a_3}^{m_1} K_{b_3,a_2}^{m_2} K_{b_4,a_1}^{n_3}
   K_{s,t}^{n_1}\\
   &+&K_{u,a_1}^{n_3} K_{b_1,a_4}^{n_4} K_{b_2,a_3}^{m_1} K_{b_3,a_2}^{m_2} K_{b_4,r}^{n_1} K_{s,t}^{n_2}+K_{u,a_4}^{n_4} K_{b_1,r}^{n_2} K_{b_2,a_3}^{m_1}
   K_{b_3,a_2}^{m_2} K_{b_4,a_1}^{n_1} K_{s,t}^{n_3}\\
   &+&K_{u,a_4}^{n_1} K_{b_1,r}^{n_3} K_{b_2,a_3}^{m_1} K_{b_3,a_2}^{m_2} K_{b_4,a_1}^{n_2}
   K_{s,t}^{n_4}+K_{s,a_1}^{n_2} K_{u,a_3}^{m_2} K_{b_1,a_2}^{n_3} K_{b_2,a_4}^{n_4} K_{b_3,t}^{m_1} K_{b_4,r}^{n_1}\\
   &+&K_{s,a_1}^{n_2} K_{u,a_3}^{m_1} K_{b_1,a_2}^{n_3}
   K_{b_2,a_4}^{n_4} K_{b_3,t}^{m_2} K_{b_4,r}^{n_1}+K_{s,a_1}^{n_2} K_{u,a_4}^{n_4} K_{b_1,t}^{n_3} K_{b_2,a_3}^{m_1} K_{b_3,a_2}^{m_2}
   K_{b_4,r}^{n_1}\\
   &+&K_{s,a_3}^{m_2} K_{u,a_1}^{n_2} K_{b_1,a_2}^{n_3} K_{b_2,a_4}^{n_4} K_{b_3,r}^{m_1} K_{b_4,t}^{n_1}+K_{s,a_3}^{m_1} K_{u,a_1}^{n_2}
   K_{b_1,a_2}^{n_3} K_{b_2,a_4}^{n_4} K_{b_3,r}^{m_2} K_{b_4,t}^{n_1}\\
   &+&K_{s,a_4}^{n_4} K_{u,a_1}^{n_2} K_{b_1,r}^{n_3} K_{b_2,a_3}^{m_1} K_{b_3,a_2}^{m_2}
   K_{b_4,t}^{n_1}+K_{s,a_1}^{n_3} K_{u,r}^{n_2} K_{b_1,a_4}^{n_4} K_{b_2,a_3}^{m_1} K_{b_3,a_2}^{m_2} K_{b_4,t}^{n_1}\\
   &+&K_{s,a_3}^{m_2} K_{u,a_4}^{n_4}
   K_{b_1,a_2}^{n_2} K_{b_2,t}^{n_3} K_{b_3,r}^{m_1} K_{b_4,a_1}^{n_1}+K_{s,a_3}^{m_2} K_{u,a_2}^{n_3} K_{b_1,t}^{n_2} K_{b_2,a_4}^{n_4} K_{b_3,r}^{m_1}
   K_{b_4,a_1}^{n_1}\\
   &+&K_{s,a_4}^{n_4} K_{u,a_3}^{m_2} K_{b_1,a_2}^{n_2} K_{b_2,r}^{n_3} K_{b_3,t}^{m_1} K_{b_4,a_1}^{n_1}+K_{s,a_2}^{n_3} K_{u,a_3}^{m_2}
   K_{b_1,r}^{n_2} K_{b_2,a_4}^{n_4} K_{b_3,t}^{m_1} K_{b_4,a_1}^{n_1}\\
   &+&K_{s,a_3}^{m_1} K_{u,a_4}^{n_4} K_{b_1,a_2}^{n_2} K_{b_2,t}^{n_3} K_{b_3,r}^{m_2}
   K_{b_4,a_1}^{n_1}+K_{s,a_3}^{m_1} K_{u,a_2}^{n_3} K_{b_1,t}^{n_2} K_{b_2,a_4}^{n_4} K_{b_3,r}^{m_2} K_{b_4,a_1}^{n_1}\\
   &+&K_{s,a_4}^{n_4} K_{u,a_3}^{m_1}
   K_{b_1,a_2}^{n_2} K_{b_2,r}^{n_3} K_{b_3,t}^{m_2} K_{b_4,a_1}^{n_1}+K_{s,a_2}^{n_3} K_{u,a_3}^{m_1} K_{b_1,r}^{n_2} K_{b_2,a_4}^{n_4} K_{b_3,t}^{m_2}
   K_{b_4,a_1}^{n_1}\\
   &+&K_{s,a_4}^{n_4} K_{u,r}^{n_3} K_{b_1,t}^{n_2} K_{b_2,a_3}^{m_1} K_{b_3,a_2}^{m_2} K_{b_4,a_1}^{n_1}+K_{s,a_1}^{n_3} K_{u,a_4}^{n_1}
   K_{b_1,t}^{n_4} K_{b_2,a_3}^{m_1} K_{b_3,a_2}^{m_2} K_{b_4,r}^{n_2}\\
   &+&K_{s,a_4}^{n_1} K_{u,a_1}^{n_3} K_{b_1,r}^{n_4} K_{b_2,a_3}^{m_1} K_{b_3,a_2}^{m_2}
   K_{b_4,t}^{n_2}+K_{s,a_3}^{m_2} K_{u,a_4}^{n_1} K_{b_1,a_2}^{n_3} K_{b_2,t}^{n_4} K_{b_3,r}^{m_1} K_{b_4,a_1}^{n_2}\\
   &+&K_{s,a_4}^{n_1} K_{u,a_3}^{m_2}
   K_{b_1,a_2}^{n_3} K_{b_2,r}^{n_4} K_{b_3,t}^{m_1} K_{b_4,a_1}^{n_2}+K_{s,a_3}^{m_1} K_{u,a_4}^{n_1} K_{b_1,a_2}^{n_3} K_{b_2,t}^{n_4} K_{b_3,r}^{m_2}
   K_{b_4,a_1}^{n_2}\\
   &+&K_{s,a_4}^{n_1} K_{u,a_3}^{m_1} K_{b_1,a_2}^{n_3} K_{b_2,r}^{n_4} K_{b_3,t}^{m_2} K_{b_4,a_1}^{n_2}+K_{s,a_4}^{n_1} K_{u,r}^{n_4} K_{b_1,t}^{n_3}
   K_{b_2,a_3}^{m_1} K_{b_3,a_2}^{m_2} K_{b_4,a_1}^{n_2}\\
   &+&\left.K_{s,a_4}^{n_2} K_{u,r}^{n_1} K_{b_1,t}^{n_4} K_{b_2,a_3}^{m_1} K_{b_3,a_2}^{m_2} K_{b_4,a_1}^{n_3}
    \right)
  \end{array}
\end{equation}
\begin{equation}
  \begin{array}{rcl}
    \Gamma_{64}^{rstu}&=& g_{6,(n_1,n_2,n_3,n_4),(m),(o)}\phi_{a_1b_1}\phi_{a_2b_2}\phi_{a_3b_3}\phi_{a_4b_4}\left(
    K_{u,a_1}^{n_2} K_{b_1,a_2}^{n_3} K_{b_2,a_4}^{n_4} K_{b_3,a_3}^o K_{b_4,t}^{n_1} K_{s,r}^m\right.\\
    &+&K_{u,a_4}^{n_4} K_{b_1,a_2}^{n_2} K_{b_2,t}^{n_3} K_{b_3,a_3}^o
   K_{b_4,a_1}^{n_1} K_{s,r}^m+K_{u,a_2}^{n_3} K_{b_1,t}^{n_2} K_{b_2,a_4}^{n_4} K_{b_3,a_3}^o K_{b_4,a_1}^{n_1} K_{s,r}^m\\
   &+&K_{u,t}^o K_{b_1,a_2}^{n_2}
   K_{b_2,a_3}^{n_3} K_{b_3,a_4}^{n_4} K_{b_4,a_1}^{n_1} K_{s,r}^m+K_{u,a_4}^{n_1} K_{b_1,a_2}^{n_3} K_{b_2,t}^{n_4} K_{b_3,a_3}^o K_{b_4,a_1}^{n_2}
   K_{s,r}^m\\
   &+&K_{u,a_1}^{n_2} K_{b_1,a_2}^{n_3} K_{b_2,a_4}^{n_4} K_{b_3,a_3}^m K_{b_4,t}^{n_1} K_{s,r}^o+K_{u,a_4}^{n_4} K_{b_1,a_2}^{n_2} K_{b_2,t}^{n_3}
   K_{b_3,a_3}^m K_{b_4,a_1}^{n_1} K_{s,r}^o\\
   &+&K_{u,a_2}^{n_3} K_{b_1,t}^{n_2} K_{b_2,a_4}^{n_4} K_{b_3,a_3}^m K_{b_4,a_1}^{n_1} K_{s,r}^o+K_{u,t}^m K_{b_1,a_2}^{n_2}
   K_{b_2,a_3}^{n_3} K_{b_3,a_4}^{n_4} K_{b_4,a_1}^{n_1} K_{s,r}^o\\
   &+&K_{u,a_4}^{n_1} K_{b_1,a_2}^{n_3} K_{b_2,t}^{n_4} K_{b_3,a_3}^m K_{b_4,a_1}^{n_2}
   K_{s,r}^o+K_{s,a_1}^{n_2} K_{u,t}^o K_{b_1,a_2}^{n_3} K_{b_2,a_4}^{n_4} K_{b_3,a_3}^m K_{b_4,r}^{n_1}\\
   &+&K_{s,a_1}^{n_2} K_{u,a_4}^{n_4} K_{b_1,t}^{n_3} K_{b_2,a_2}^m
   K_{b_3,a_3}^o K_{b_4,r}^{n_1}+K_{s,t}^{n_2} K_{u,a_1}^{n_3} K_{b_1,a_4}^{n_4} K_{b_2,a_2}^m K_{b_3,a_3}^o K_{b_4,r}^{n_1}\\
   &+&K_{s,a_1}^{n_2} K_{u,t}^m
   K_{b_1,a_2}^{n_3} K_{b_2,a_4}^{n_4} K_{b_3,a_3}^o K_{b_4,r}^{n_1}+K_{s,a_4}^{n_4} K_{u,a_1}^{n_2} K_{b_1,r}^{n_3} K_{b_2,a_2}^m K_{b_3,a_3}^o
   K_{b_4,t}^{n_1}\\
   &+&K_{s,a_1}^{n_3} K_{u,r}^{n_2} K_{b_1,a_4}^{n_4} K_{b_2,a_2}^m K_{b_3,a_3}^o K_{b_4,t}^{n_1}+K_{s,a_4}^{n_4} K_{u,t}^o K_{b_1,a_2}^{n_2}
   K_{b_2,r}^{n_3} K_{b_3,a_3}^m K_{b_4,a_1}^{n_1}\\
   &+&K_{s,a_2}^{n_3} K_{u,t}^o K_{b_1,r}^{n_2} K_{b_2,a_4}^{n_4} K_{b_3,a_3}^m K_{b_4,a_1}^{n_1}+K_{s,t}^{n_3}
   K_{u,a_4}^{n_4} K_{b_1,r}^{n_2} K_{b_2,a_2}^m K_{b_3,a_3}^o K_{b_4,a_1}^{n_1}\\
   &+&K_{s,a_4}^{n_4} K_{u,r}^{n_3} K_{b_1,t}^{n_2} K_{b_2,a_2}^m K_{b_3,a_3}^o
   K_{b_4,a_1}^{n_1}+K_{s,a_4}^{n_4} K_{u,t}^m K_{b_1,a_2}^{n_2} K_{b_2,r}^{n_3} K_{b_3,a_3}^o K_{b_4,a_1}^{n_1}\\
   &+&K_{s,a_2}^{n_3} K_{u,t}^m K_{b_1,r}^{n_2}
   K_{b_2,a_4}^{n_4} K_{b_3,a_3}^o K_{b_4,a_1}^{n_1}+K_{s,a_1}^{n_3} K_{u,a_4}^{n_1} K_{b_1,t}^{n_4} K_{b_2,a_2}^m K_{b_3,a_3}^o K_{b_4,r}^{n_2}\\
   &+&K_{s,a_4}^{n_1}
   K_{u,a_1}^{n_3} K_{b_1,r}^{n_4} K_{b_2,a_2}^m K_{b_3,a_3}^o K_{b_4,t}^{n_2}+K_{s,a_4}^{n_1} K_{u,t}^o K_{b_1,a_2}^{n_3} K_{b_2,r}^{n_4} K_{b_3,a_3}^m
   K_{b_4,a_1}^{n_2}\\
   &+&K_{s,t}^{n_4} K_{u,a_4}^{n_1} K_{b_1,r}^{n_3} K_{b_2,a_2}^m K_{b_3,a_3}^o K_{b_4,a_1}^{n_2}+K_{s,a_4}^{n_1} K_{u,r}^{n_4} K_{b_1,t}^{n_3}
   K_{b_2,a_2}^m K_{b_3,a_3}^o K_{b_4,a_1}^{n_2}\\
   &+&K_{s,a_4}^{n_1} K_{u,t}^m K_{b_1,a_2}^{n_3} K_{b_2,r}^{n_4} K_{b_3,a_3}^o K_{b_4,a_1}^{n_2}+K_{s,t}^{n_1}
   K_{u,a_4}^{n_2} K_{b_1,r}^{n_4} K_{b_2,a_2}^m K_{b_3,a_3}^o K_{b_4,a_1}^{n_3}\\
   &+&\left.K_{s,a_4}^{n_2} K_{u,r}^{n_1} K_{b_1,t}^{n_4} K_{b_2,a_2}^m K_{b_3,a_3}^o
   K_{b_4,a_1}^{n_3}
    \right).
  \end{array}
\end{equation}
Let us now investigate the r.h.s. of the flow equation for the individual couplings in the self-dual GW-action.

{\it The r.h.s. for $m^2$} is the restriction $\left.-\frac 1 2 \dot R^{abcd}P_{cdrs}(\Gamma_4[\phi]-R_4[\phi])^{rstu}P_{tuab}\right|_{[\phi.\phi]_k}$. Let us denote the index independent part of $\frac 1 2 (P.\dot R.P)^{rstu}$ by $I^{rstu}$ and its index dependent part by $D^{rstu}$, such that $[\phi.\phi]_k$ is generated by the following terms on the r.h.s. of the flow equation
\begin{equation}
  \begin{array}{l}
    (D+I)^{rstu}[\phi.\phi]A^{rstu}\\
    I^{rstu}\left((\phi.\phi)^{ru}A^{st}+...\right),
  \end{array}
\end{equation}
where $A$ stands for any rest and the parenthesis stands for permutation of indices.

{\it The r.h.s. for $Z$} is $\left.-\frac 1 2 \dot R^{abcd}P_{cdrs}(\Gamma_4[\phi]-R_4[\phi])^{rstu}P_{tuab}\right|_{[\phi.K.\phi]_k}$, which is generated by the following terms on the r.h.s. of the flow equation:
\begin{equation}
  \begin{array}{l}
    (D+I)^{rstu}\left([\phi.\phi.K]+[\phi.K.\phi]+[\phi.\phi.K]\right)A^{rstu}\\
    D^{rstu}\left((\phi.\phi)^{st}A^{ru}+...\right)\\
    I^{rstu}\left(\left(K.\phi.\phi+\phi.K.\phi+\phi.\phi.K\right)^{st}A^{ru}+...\right).
  \end{array}
\end{equation}

{\it The r.h.s. for $\lambda$} is $\left.-\frac 1 2 \dot R^{abcd}P_{cdrs}(\Gamma_6[\phi]-R_6[\phi])^{rstu}P_{tuab}\right|_{[\phi.\phi.\phi.\phi]_k} +$ \\ $\left.\frac 1 2 \dot R^{abcd}P_{cdr_1s_1}(\Gamma_4[\phi]-R_4[\phi])^{r_1s_1t_1u_1}P^{t_1u_1r_2s_2}(\Gamma_4[\phi]-R_4[\phi])^{r_2s_2t_2u_2}P_{t_2u_2ab}\right|_{[\phi.\phi.\phi.\phi]_k}$,
which is generated by the following terms on the r.h.s. of the flow equation:
\begin{equation}
  \begin{array}{l}
    (I+D)^{rstu}[\phi.\phi.\phi.\phi]A^{rstu}\\
    I^{rstu}\left((\phi.\phi.\phi.\phi)^{st}A^{ru}+...\right)\\
    I^{r_1s_1t_2u_2}\left((\phi.\phi)^{s_1t_1} A^{r_1u_1}+(\phi.\phi)^{s_1t_1} A^{r_1u_1}\right)C\delta^{t_1u_1r_2s_2}\left((\phi.\phi)^{s_2t_2} A^{r_2u_2}+(\phi.\phi)^{s_2t_2} A^{r_2u_2}\right)
  \end{array}
\end{equation}
where we split $P^{rstu}=C\delta^{ru}\delta^{st}+Q^{rstu}$ into its index independent diagonal part and rest. This convenient form of the propagator comes straight from the inversion of $\Gamma^{(2)}+R$, because $R$ is supposed to give IR modes a large mass term $O(k^2) \delta^{ru}\delta^{st}$, such that we can invert $\Gamma^{(2)}+R$ in a geometric series $C^{-1}(\mathbb I+C^{-1}\textrm{rest})^{-1}=C^{-1} \mathbb I-C^{-2}\textrm{rest}+...$ ,where we used that $\delta^{ru}\delta^{st}$ is the identity operator $\mathbb I$ on matrices.

Let us now extract the necessary terms $(\phi.\phi)^{st}A^{ru},(\phi.\phi)^{ru}A^{st},[\phi.\phi]A^{rstu},...$ from the Hessains. Although possible it is very cumbersome to discuss all terms individually. We thus automatize this procedure by defining a set of contraction rules that yield the desired term and subsequent extraction of the coefficient of this term.

To extract $(\phi.\phi)^{..},[\phi.\phi]$ we apply to $(\Gamma_4[\phi]-R_4[\phi])^{rstu}$ the repeated replacement rules
\begin{equation}\label{rep:m2-extract}
  \begin{array}{rcl}
    K_{ab}^n\phi_{bc}&\to&\delta_{n,0} \phi_{ac}\\
    K^n_{ba}\phi_{bc}&\to&\delta_{n,0} \phi_{ac}\\
    K_{ab}^n\phi_{cb}&\to&\delta_{n,0} \phi_{ca}\\
    K^n_{ba}\phi_{cb}&\to&\delta_{n,0} \phi_{ca}\\
    \phi_{ab}\phi_{bc}&\to&\sigma \phi.\phi_{ac},
  \end{array}
\end{equation}
and subsequently extract the coefficient of $\sigma$ and investigate the index structure of this coefficient. This yields the coefficients
\begin{equation}
  \begin{array}{rcl}
    (\phi.\phi)^{sr}&:&3  K_{u,t}^m g_{4,(0,0,0),(m)}\\
    (\phi.\phi)^{st}&:&  g_{4,({n_1},0,0,0)} K_{u,r}^{n_1}+  g_{4,(0,{n_2},0,0)} K_{u,r}^{n_2}+  g_{4,(0,0,{n_3},0)} K_{u,r}^{n_3}+  g_{4,(0,0,0,{n_4})}
   K_{u,r}^{n_4}\\
    (\phi.\phi)^{ur}&:&  g_{4,({n_1},0,0,0)} K_{s,t}^{n_1}+  g_{4,(0,{n_2},0,0)} K_{s,t}^{n_2}+  g_{4,(0,0,{n_3},0)} K_{s,t}^{n_3}+  g_{4,(0,0,0,{n_4})}
   K_{s,t}^{n_4}\\
    (\phi.\phi)^{ut}&:&3  K_{s,r}^m g_{4,(0,0,0),(m)}\\
    {}[ {\phi.\phi} ]   &:& K_{u,t}^o g_{4,(0,0),(m),(o)} K_{s,r}^m+ K_{u,t}^m g_{4,(0,0),(m),(o)} K_{s,r}^o+ K_{s,t}^{m_2} K_{u,r}^{m_1}
   g_{4,(0,0),({m_1},{m_2})}\\
   &+& K_{s,t}^{m_1} K_{u,r}^{m_2} g_{4,(0,0),({m_1},{m_2})}+ K_{s,t}^{n_2} K_{u,r}^{n_1}
   g_{4,({n_1},{n_2}),(0,0)}+ K_{s,t}^{n_1} K_{u,r}^{n_2} g_{4,({n_1},{n_2}),(0,0)}
  \end{array}
\end{equation}
imposing the uniqueness conditions for the couplings yields
\begin{equation}
  \begin{array}{rcl}
    (\phi.\phi)^{sr}&:&3  K_{u,t}^m g_{4,(0,0,0),(m)}\\
    (\phi.\phi)^{st}&:&  g_{4,({n_1},0,0,0)} K_{u,r}^{n_1}+3  g_{4,(0,0,0,0)} \delta_{u,r}\\
    (\phi.\phi)^{ur}&:&  g_{4,({n_1},0,0,0)} K_{s,t}^{n_1}+3  g_{4,(0,0,0,0)} \delta_{s,t}\\
    (\phi.\phi)^{ut}&:&3  K_{s,r}^m g_{4,(0,0,0),(m)}\\
    {}[ {\phi.\phi} ]   &:& g_{4,(0,0),(m),(o)} \left(K_{u,t}^o  K_{s,r}^m+K_{u,t}^m K_{s,r}^o\right)+2
   g_{4,(0,0),0,0)}\delta_{s,t} \delta_{u,r}\\
   &+&
   g_{4,(\text{n1},{n2}),(0,0)}\left(K_{s,t}^{n_2} K_{u,r}^{n_1}+ K_{s,t}^{n_1} K_{u,r}^{n_2}\right).
  \end{array}
\end{equation}

To extract $(K.\phi.\phi)^{..},(\phi.K.\phi)^{..},(\phi.\phi.K)^{..},[K.\phi.\phi]$, we apply to $(\Gamma_4[\phi]-R_4[\phi])^{rstu}$ the repeated replacements
\begin{equation}
  \begin{array}{rcl}
    K^n_{ab}\phi_{bc}&\to&\delta_{n,0} \phi_{ac}+\delta_{n,1} K.\phi_{ac}\\
    \phi_{ab}K^n_{bc}&\to&\delta_{n,0} \phi_{ac}+\delta_{n,1} \phi.K_{ac}\\
    \phi_{ab}\phi_{bc}&\to&\phi.\phi_{ac}\\
    K^n_{ab}\phi.\phi_{bc}&\to&\sigma \delta_{n,1}K.\phi.\phi_{ac}\\
    \phi.\phi_{ab}K^n_{bc}&\to&\sigma \delta_{n,1}\phi.\phi.K_{ac}\\
    \phi_{ab}K.\phi_{bc}&\to&\sigma \phi.K.\phi_{ac}\\
    K.\phi_{ab}\phi_{bc}&\to&\sigma K.\phi.\phi_{ac}\\
    \phi_{ab}\phi.K_{bc}&\to&\sigma \phi.\phi.K_{ac}\\
    \phi.K_{ab}\phi_{bc}&\to&\sigma \phi.K.\phi_{ac}
  \end{array}
\end{equation}
and subsequently extract the coefficient of $\sigma$. Comparing the index structure yields the coefficients
\begin{equation}
  \begin{array}{rcl}
    (K.\phi.\phi)^{ur}&:&  g_{4,({n_1},1,0,0)} K_{s,t}^{n_1}+  g_{4,(0,{n_2},1,0)} K_{s,t}^{n_2}+  g_{4,(0,0,{n_3},1)} K_{s,t}^{n_3}+  g_{4,(1,0,0,{n_4})}
   K_{s,t}^{n_4}\\
    (K.\phi.\phi)^{st}&:&  g_{4,({n_1},1,0,0)} K_{u,r}^{n_1}+  g_{4,(0,{n_2},1,0)} K_{u,r}^{n_2}+  g_{4,(0,0,{n_3},1)} K_{u,r}^{n_3}+  g_{4,(1,0,0,{n_4})}
   K_{u,r}^{n_4}\\
    (K.\phi.\phi)^{ut}&:& g_{4,(0,0,1),(m)} K_{s,r}^m+ g_{4,(0,1,0),(m)} K_{s,r}^m+ g_{4,(1,0,0),(m)} K_{s,r}^m\\
    (K.\phi.\phi)^{sr}&:& g_{4,(0,0,1),(m)} K_{u,t}^m+ g_{4,(0,1,0),(m)} K_{u,t}^m+ g_{4,(1,0,0),(m)} K_{u,t}^m\\
    {}[K.\phi.\phi]&:& K_{u,r}^{m_2} \left(g_{4,(0,1),({m_1},{m_2})}+g_{4,(1,0),({m_1},{m_2})}\right) K_{s,t}^{m_1}\\
    &+& K_{u,r}^{m_1}
   \left(g_{4,(0,1),({m_1},{m_2})}+g_{4,(1,0),({m_1},{m_2})}\right) K_{s,t}^{m_2}\\
   &+& K_{u,r}^{n_2}
   \left(g_{4,({n_1},{n_2}),(0,1)}+g_{4,({n_1},{n_2}),(1,0)}\right) K_{s,t}^{n_1}\\
   &+& K_{u,r}^{n_1}
   \left(g_{4,({n_1},{n_2}),(0,1)}+g_{4,({n_1},{n_2}),(1,0)}\right) K_{s,t}^{n_2}\\
   &+& \left(K_{u,t}^o K_{s,r}^m+K_{u,t}^m K_{s,r}^o\right)
   \left(g_{4,(0,1),(m),(o)}+g_{4,(1,0),(m),(o)}\right)\\
    (\phi.\phi.K)^{ur}&:&  g_{4,({n_1},0,0,1)} K_{s,t}^{n_1}+  g_{4,(1,{n_2},0,0)} K_{s,t}^{n_2}+  g_{4,(0,1,{n_3},0)} K_{s,t}^{n_3}+  g_{4,(0,0,1,{n_4})}
   K_{s,t}^{n_4}\\
    (\phi.\phi.K)^{st}&:&  g_{4,({n_1},0,0,1)} K_{u,r}^{n_1}+  g_{4,(1,{n_2},0,0)} K_{u,r}^{n_2}+  g_{4,(0,1,{n_3},0)} K_{u,r}^{n_3}+  g_{4,(0,0,1,{n_4})}
   K_{u,r}^{n_4}\\
    (\phi.\phi.K)^{ut}&:& g_{4,(0,0,1),(m)} K_{s,r}^m+ g_{4,(0,1,0),(m)} K_{s,r}^m+ g_{4,(1,0,0),(m)} K_{s,r}^m\\
    (\phi.\phi.K)^{sr}&:& g_{4,(0,0,1),(m)} K_{u,t}^m+ g_{4,(0,1,0),(m)} K_{u,t}^m+ g_{4,(1,0,0),(m)} K_{u,t}^m\\
  \end{array}
\end{equation}
where the index uniqueness conditions yield:
\begin{equation}
  \begin{array}{rcl}
    (K.\phi.\phi)^{ur}&:&  g_{4,({n_1},1,0,0)} K_{s,t}^{n_1}+  g_{4,(1,0,0,0)}
   \delta_{s,t}+ g_{4,(1,0,0,1)}K_{s,t}\\
    (K.\phi.\phi)^{st}&:&  g_{4,({n_1},1,0,0)} K_{u,r}^{n_1}+  g_{4,(1,0,0,0)}
   \delta_{u,r}+ g_{4(1,0,0,1)}K_{u,r}\\
    (K.\phi.\phi)^{ut}&:& g_{4,(1,0,0),(m)} K_{s,r}^m\\
    (K.\phi.\phi)^{sr}&:& g_{4,(1,0,0),(m)} K_{u,t}^m\\
    {}[K.\phi.\phi]&:&2\left(g_{4,(1,0)(0,0)}\delta_{s,t}\delta_{u,r}+g_{4,(1,0)(1,1)}K_{s,t}K_{u,r}\right)\\
    &+& g_{4,(1,0)(1,0)}\left(\delta_{s,t}K_{u,r}+\delta_{u,r}K_{s,t}\right)\\
   &+&  g_{4,({n_1},{n_2}),(1,0)} \left(K_{s,t}^{n_2}K_{u,r}^{n_1}+K_{s,t}^{n_1}K_{u,r}^{n_2}\right)\\
   &+& \left(K_{u,t}^o K_{s,r}^m+K_{u,t}^m K_{s,r}^o\right)g_{4,(1,0),(m),(o)}\\
    (\phi.\phi.K)^{ur}&:&  g_{4,({n_1},0,0,1)} K_{s,t}^{n_1}+ \left( g_{4,(1,0,0,0)} \delta_{s,t}+g_{4,(1,1,0,0)} K_{s,t}\right)\\
    (\phi.\phi.K)^{st}&:&  g_{4,({n_1},0,0,1)} K_{u,r}^{n_1}+  \left(g_{4,(1,0,0,0)} \delta_{u,r}+g_{4,(1,1,0,0)} K_{u,r}\right)\\
    (\phi.\phi.K)^{ut}&:& g_{4,(1,0,0),(m)} K_{s,r}^m\\
    (\phi.\phi.K)^{sr}&:& g_{4,(1,0,0),(m)} K_{u,t}^m\\
  \end{array}
\end{equation}

To extract $[\phi.\phi.\phi.\phi],(\phi.\phi.\phi.\phi)^{..}$ from $(\Gamma_6[\phi]-R_6[\phi])^{rstu}$, we use repeated replacements analogous to (\ref{rep:m2-extract}). The coefficients turn out to be
\begin{equation}
  \begin{array}{rcl}
    (\phi.\phi.\phi.\phi)^{st}&:&  g_{6,({n_1},0,0,0,0,0)} K_{u,r}^{n_1}+  g_{6,(0,{n_2},0,0,0,0)} K_{u,r}^{n_2}+  g_{6,(0,0,{n_3},0,0,0)} K_{u,r}^{n_3}\\
    &+&
   g_{6,(0,0,0,{n_4},0,0)} K_{u,r}^{n_4}+  g_{6,(0,0,0,0,{n_5},0)} K_{u,r}^{n_5}+  g_{6,(0,0,0,0,0,{n_6})} K_{u,r}^{n_6}\\
    (\phi.\phi.\phi.\phi)^{ur}&:&  g_{6,({n_1},0,0,0,0,0)} K_{s,t}^{n_1}+  g_{6,(0,{n_2},0,0,0,0)} K_{s,t}^{n_2}+  g_{6,(0,0,{n_3},0,0,0)} K_{s,t}^{n_3}\\
    &+&
   g_{6,(0,0,0,{n_4},0,0)} K_{s,t}^{n_4}+  g_{6,(0,0,0,0,{n_5},0)} K_{s,t}^{n_5}+  g_{6,(0,0,0,0,0,{n_6})} K_{s,t}^{n_6}\\
    (\phi.\phi.\phi.\phi)^{sr}&:&5  K_{u,t}^m g_{6,(0,0,0,0,0),(m)}\\
    (\phi.\phi.\phi.\phi)^{ut}&:&5  K_{s,r}^m g_{6,(0,0,0,0,0),(m)}\\
    {}[\phi.\phi.\phi.\phi]   &:& K_{u,t}^o g_{6,(0,0,0,0),(m),(o)} K_{s,r}^m+ K_{u,t}^m g_{6,(0,0,0,0),(m),(o)} K_{s,r}^o\\
    &+& K_{s,t}^{m_2} K_{u,r}^{m_1}
   g_{6,(0,0,0,0),({m}_1,{m}_2)}+ K_{s,t}^{m_1} K_{u,r}^{m_2} g_{6,(0,0,0,0),({m_1},{m_2})}.
  \end{array}
\end{equation}
Imposing the uniqueness conditions for the couplings yields:
\begin{equation}
  \begin{array}{rcl}
    (\phi.\phi.\phi.\phi)^{st}&:&  g_{6,({n_1},0,0,0,0,0)} K_{u,r}^{n_1}+5  g_{6,(0,0,0,0,0,0)} \delta_{u,r}\\
    (\phi.\phi.\phi.\phi)^{ur}&:&  g_{6,({n_1},0,0,0,0,0)} K_{s,t}^{n_1}+5  g_{6,(0,0,0,0,0,0)} \delta_{s,t}\\
    (\phi.\phi.\phi.\phi)^{sr}&:&5  K_{u,t}^m g_{6,(0,0,0,0,0),(m)}\\
    (\phi.\phi.\phi.\phi)^{ut}&:&5  K_{s,r}^m g_{6,(0,0,0,0,0),(m)}\\
    {}[\phi.\phi.\phi.\phi]   &:& g_{6,(0,0,0,0),(m),(o)}\left( K_{u,t}^o  K_{s,r}^m+ K_{u,t}^m  K_{s,r}^o\right)\\
    &+&
   g_{6,(0,0,0,0),({m}_1,{m}_2)}\left(K_{s,t}^{m_2} K_{u,r}^{m_1}+ K_{s,t}^{m_1} K_{u,r}^{m_2}\right).
  \end{array}
\end{equation}

\end{adjustwidth}

\end{document}